\documentclass[superscriptaddress,preprintnumbers,nofootinbib,showpacs,twocolumn]{revtex4}

\usepackage{graphicx}
\usepackage{latexsym}
\usepackage[english]{babel}
\usepackage{amsmath}
\usepackage{amsthm}
\usepackage{epsfig}
\usepackage{color}
\usepackage{amssymb}

\begin{document}

\title{\Large Fermion Propagator in Cosmological Spaces with Constant Deceleration}

\preprint{ITP-UU-09/01, SPIN-09/01}


\author{Jurjen F. Koksma}
\email[]{J.F.Koksma@uu.nl} \affiliation{Institute for Theoretical
Physics (ITP) \& Spinoza Institute, Utrecht University, Postbus
80195, 3508 TD Utrecht, The Netherlands}

\author{Tomislav Prokopec}
\email[]{T.Prokopec@uu.nl} \affiliation{Institute for Theoretical
Physics (ITP) \& Spinoza Institute, Utrecht University, Postbus
80195, 3508 TD Utrecht, The Netherlands}

\begin{abstract}
We calculate the fermion propagator in FLRW spacetimes with
constant deceleration $q=\epsilon-1$, $\epsilon=-\dot{H}/H^{2}$
for excited states. For fermions whose mass is generated by a
scalar field through a Yukawa coupling
$m=g_{\mathrm{\scriptscriptstyle{Y}}} \phi$, we assume $\phi
\propto H$. We first solve for the mode functions by splitting the
spinor into a direct product of helicity and chirality spinors. We
also allow for non-vacuum states. We normalise the spinors using a
consistent canonical quantisation and by requiring orthogonality
of particle and anti-particle spinors. We apply our propagator to
calculate the one loop effective action and renormalise using
dimensional regularisation. Since the Hubble parameter is now
treated dynamically, this paves the way to study the dynamical
backreaction of fermions on the background spacetime.
\end{abstract}

\maketitle

\section{Introduction}
\label{Introduction}

Partly motivated by our increased capability in recent years to
perform accurate observations on the sky, we have turned our
attention to investigating the impact of quantum effects on the
evolution of the universe. In pursuit of this goal, quantum field
theory in cosmological spacetimes continues to be an increasingly
important field of research. Among other things, it investigates
how quantum fluctuations affect the background spacetime in
perturbative quantum gravity, a process also known as quantum
backreaction.

A line of research deals for example with the backreaction of
quantum fields whose spectrum is nearly flat. Examples of these
fields are the minimally coupled massless scalar and the graviton.
Consequently, these fields are expected to yield a substantial
backreaction \cite{Ford:1984hs, Antoniadis:1986sb, Tsamis:1992xa,
Tsamis:1996qq, Tsamis:1996qk, Abramo:1997hu, Finelli:2004bm,
Prokopec:2006yh, Prokopec:2007ak, Bilandzic:2007nb,
Janssen:2007ht, Janssen:2008dp, Janssen:2008dw,Janssen:2008px}.
Also, fermions in curved spacetimes and in particular maximally
symmetric spacetimes were investigated (see e.g.:
\cite{Candelas:1975du, Allen:1986qj, Camporesi:1992tm,
Miao:2005am, Miao:2006pn}).

Another approach to this question is concerned with calculating
the possible effect of the trace anomaly on the background
spacetime. Some authors argue that this effect, particularly in
relation with the cosmological constant problem, could be
significant \cite{Tomboulis:1988gw, Antoniadis:1991fa,
Antoniadis:1992hz, Antoniadis:1998fi, Salehi:2000eu,
Antoniadis:2006wq}, however also see \cite{Koksma:2008jn}.

An essential element to study quantum effects in curved spaces is
the propagator. Due to its high degree of symmetry, de Sitter
spacetime proves to be ideally suited for calculating various
quantum effects. Moreover, results are immediately applicable to
inflation, cosmologists' favourite paradigm for a brief
exponentially fast expansion in the early universe. Chernikov and
Tagirov calculated the scalar propagator in de Sitter spacetime
\cite{Chernikov:1968zm}, also see \cite{Allen:1985ux,
Allen:1987tz}. The vector propagator was constructed by
\cite{Allen:1985wd, Tsamis:2006gj} and the graviton propagator
received contributions from \cite{Allen:1986ta, Allen:1986tt,
Antoniadis:1986sb, Floratos:1987ek, Tsamis:1992xa,
Higuchi:2001uv}. Candelas and Raine calculated the fermion
propagator \cite{Candelas:1975du}.

However, de Sitter spacetime suffers from several drawbacks.
Firstly, pure de Sitter spacetime is never realised in nature.
Since de Sitter spacetime corresponds to a globally constant
Hubble parameter $H$, this is in reality never attained. Secondly,
de Sitter spacetime is non-dynamical. It is therefore inconsistent
to study backreaction effects while at the same time assuming a
constant background. We should allow the background to change to
encompass all possible backreaction effects consistently.
Therefore, it is necessary to consider propagators in more general
spacetimes.

The scalar and graviton propagator in quasi de Sitter spacetime
were calculated by \cite{Janssen:2007ht} and subsequently
generalised to Friedmann-Lema\^itre-Robertson-Walker or FLRW
spacetimes with constant deceleration by \cite{Janssen:2008dp,
Janssen:2008dw,Janssen:2008px}. In this paper we calculate the
fermion propagator in FLRW spacetimes with constant deceleration.

We should touch upon an important issue. In the massive case we
need two additional constraints to be satisfied in order to solve
for the propagator:
\begin{subequations}
\label{assumptions}
\begin{eqnarray}
\epsilon \equiv - \frac{\dot{H}}{H^{2}} &=& \mathrm{const}
\label{epsilon} \\
\frac{m}{H} &=& \mathrm{const} \label{mHassumption} \,.
\end{eqnarray}
\end{subequations}
Recall that all interesting epochs in the evolution of our
universe satisfy the first constraint. In the matter era we have
$\epsilon=3/2$, in the radiation dominated epoch we find
$\epsilon=2$. It also serves as an approximation to both inflation
and the current dark energy dominated epoch when $\epsilon \ll 1$.
Note that $\epsilon$ coincides with the slow-roll parameter of
inflation and it is straightforwardly related to the somewhat more
familiar deceleration parameter $q = \epsilon -1$. Hence
(\ref{epsilon}) is equivalent to requiring a constant deceleration
parameter.

The second assumption (\ref{mHassumption}) is required to
analytically derive the propagator in the massive case. In Yukawa
theory, the mass of the fermion is generated by a scalar field for
which we can achieve $\phi \propto H$ in several cases to which we
will turn shortly.

We then calculate the one loop effective action induced by
fermions using our propagator. The one loop backreaction arises
from integrating out a free, quadratic fermion field and, using
dimensional regularisation, this generates a correction to the
(classical) Friedmann equations.

The Hubble parameter occurring in this effective action is now a
dynamical quantity. Consequently, the backreaction of these
fermions on the background spacetime can thus be analysed
dynamically. Also, when the fermions are coupled to a scalar
field, this opens up the possibility to study the impact of the
fermions on the evolution of scalar fields.

This paper is outlined as follows. In section \ref{Fermions in
FLRW spacetimes} we review the basic theory required to study
fermions in curved spaces and we establish our notation. Sections
\ref{FLRW Fermion propagator: The Massless Case} and \ref{FLRW
Fermion propagator: The Massive Case} are devoted to deriving the
fermionic propagator in FLRW spacetimes in the massless and
massive case, respectively. Finally, in section \ref{One Loop
Effective Action} we apply our propagator to calculate the one
loop effective action.

\section{Fermions in FLRW spacetimes}
\label{Fermions in FLRW spacetimes}

\subsection{The Dirac Equation}
\label{The Dirac Equation}

Fermions are in general $D$ dimensional curved spacetimes
described by the action \cite{Birrell:1982ix}:
\begin{equation}\label{fermionaction}
S = \int \mathrm{d}^{\scriptscriptstyle{D}}\! x
\sqrt{-g}\left\{\frac{i}{2}\left[\bar{\psi} \gamma^{\mu}
\nabla_{\mu} \psi - \left (\nabla_{\mu} \bar{\psi}\right)
\gamma^{\mu} \psi\right] - m \bar{\psi}\psi\right\} ,
\end{equation}
where the Dirac matrices $\gamma^{\mu}$ satisfy the following
anti-commutation relations:
\begin{equation}\label{gammamatrices}
\left\{\gamma^{\mu},\gamma^{\nu} \right\} = - 2 g^{\mu\nu} \,.
\end{equation}
Variation with respect to $\bar{\psi} = \psi^{\dag} \gamma^{0}$
yields the equation of motion the fermion field $\psi$ satisfies:
\begin{equation}\label{fermioneom}
i \gamma^{\mu} \nabla_{\mu} \psi(x) - m \psi(x) = 0 \,.
\end{equation}
We will make use of the vierbein formalism which can be thought of
as a transformation of the metric tensor to a locally flat
Minkowski metric:
\begin{equation}\label{vierbein}
g_{\mu\nu}(x) = e_{\mu}^{a}(x) e_{\nu}^{b}(x) \eta_{a b}  \,,
\end{equation}
where $\eta_{ab} = \mathrm{diag}(-1,1,\cdots,1)$ is the Minkowski
metric. We use Greek letters to run over spacetime indices whereas
we use Latin letters either for the tangent space indices or for
spinor indices. It will be clear from the context which
interpretation we are using. The Minkowski metric and the (flat
space) Dirac matrices $\gamma^{a}$ are spacetime independent.
Spacetime and tangent space indices are raised or lowered by
making use of the full metric or Minkowski metric, respectively.
The covariant derivative acting on a Dirac spinor is defined as:
\begin{equation}\label{covariantderivative}
\nabla_{\mu} \psi(x) = \partial_{\mu} \psi(x) -
\Gamma_{\mu}\psi(x) \,,
\end{equation}
where the spin connection is given by:
\begin{equation}\label{spinconnection}
\Gamma_{\mu} = -\frac{1}{8} e^{\nu}_{c}\left( \partial_{\mu}
e_{\nu d} - \Gamma^{\alpha}_{\phantom{\alpha}\mu\nu} e_{\alpha d}
\right) \left[\gamma^{c}, \gamma^{d} \right] \,,
\end{equation}
such that $\nabla_{\mu}\gamma_{\nu} = 0$. Let us specialise to
flat FLRW spacetimes in which the metric is given by $g_{\mu\nu}=
a^{2}(\eta) \eta_{\mu\nu}$. Here, $a(\eta)$ is the scale factor of
the universe in conformal time defined by $dt = a(\eta) d\eta$. In
FLRW spacetimes the vierbeins are a function of conformal time
only:
\begin{subequations}
\begin{eqnarray}
e^{b}_{\mu}(\eta) &=& \delta^{b}_{\mu} a(\eta)\\
e_{b}^{\mu}(\eta) &=& \delta_{b}^{\mu} a^{-1}(\eta) \,.
\end{eqnarray}
\end{subequations}
We thus find:
\begin{equation}\label{covariantderivative2}
i \gamma^{\mu} \nabla_{\mu} \psi(x) = a^{-\frac{D+1}{2}}(\eta) i
\gamma^{b}\partial_{b}\left(a^{\frac{D-1}{2}}(\eta) \psi(x)
\right) \,.
\end{equation}
This useful identity relates covariant derivatives to partial
derivatives in the tangent space.

Since we work in the chirality representation, the flat space
Dirac matrices are given by:
\begin{equation}\label{gammamatrices2}
\gamma^{0} =
\begin{pmatrix}
0 & \mathbb{I}\\
\mathbb{I} & 0
\end{pmatrix}
\,,
\end{equation}
where $\mathbb{I}$ is the $2\times 2$ identity matrix and:
\begin{equation}\label{gammamatrices3}
\gamma^{i} =
\begin{pmatrix}
0 & \sigma^{i}\\
-\sigma^{i} & 0
\end{pmatrix}
\,,
\end{equation}
and where the $\sigma^{i}$ denote the Pauli matrices.
\begin{figure}[t]
 \centering
  \begin{minipage}[t]{.95\columnwidth}
   \includegraphics[width=\columnwidth]{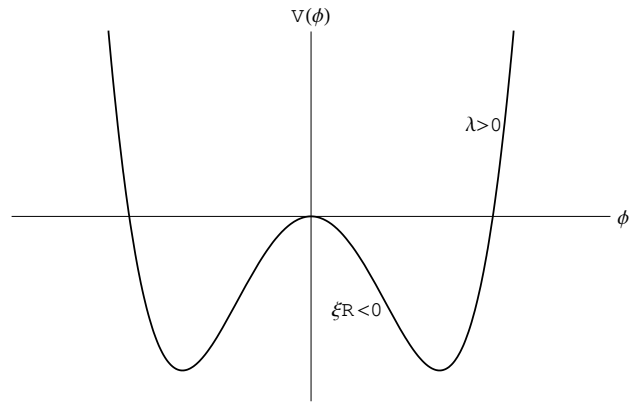}
   {\em \caption{Qualitative sketch for the ``Mexican hat'' potential
   (\ref{potential}) assuming $\mu^{2} \ll | \xi R |$. Near the
   minimum of the potential
   (\ref{phiproptoH}), assumption (\ref{mHassumption}) is
   satisfied.
   \label{fig:potential} }}
   \end{minipage}
\end{figure}

In Yukawa theory the mass of the fermion appearing in equation
(\ref{fermionaction}) is time dependent and generated by a scalar
field. Let us thus briefly consider the complex system of a
fermion whose mass is generated by a homogeneous scalar field
$\phi(\eta)$, both of which are coupled to gravity. The action is:
\begin{eqnarray}
S &=& \int \mathrm{d}^{\scriptscriptstyle{D}}\! x \sqrt{-g}\left\{
-\frac{1}{2}g^{\mu\nu}\partial_{\mu} \phi \partial_{\nu} \phi
-V(\phi)- g_{\mathrm{\scriptscriptstyle{Y}}} \bar{\psi}\psi\phi \right\} \nonumber \\
&& + \int \mathrm{d}^{\scriptscriptstyle{D}}\! x
\sqrt{-g}\left\{\frac{i}{2}\left[\bar{\psi} \gamma^{\mu}
\nabla_{\mu} \psi - \left (\nabla_{\mu} \bar{\psi}\right)
\gamma^{\mu} \psi\right] \right\}\label{actiontotal} . \phantom{1}
\end{eqnarray}
Here $g_{\mathrm{\scriptscriptstyle{Y}}}$ the Yukawa coupling
constant. By comparing equations (\ref{actiontotal}) and
(\ref{fermionaction}) we can thus identify the mass of the fermion
as:
\begin{equation}\label{fermionmass}
m(\eta) = g_{\mathrm{\scriptscriptstyle{Y}}} \phi(\eta) \,.
\end{equation}
Furthermore, the (classical) evolution of the homogeneous
background field $\phi(\eta)$ is governed by:
\begin{equation}\label{eomphi}
\Box \phi - \partial_{\phi}V(\phi, R) = 0 \,,
\end{equation}
where $V(\phi, R)$ is the potential:
\begin{equation}\label{potential}
V(\phi, R) = \frac{\mu^{2}}{2}\phi^{2} + \frac{\xi}{2}R\phi^{2} +
\frac{\lambda}{4!}\phi^{4} \,.
\end{equation}
Here, $\xi$ denotes some non-minimal coupling to gravity and $R$
is the Ricci scalar. In FLRW spacetimes in $D=4$ we have
$R=6(\dot{H}+2H^{2})= 6(2-\epsilon)H^{2}$, where the Hubble
parameter is given by $H=\dot{a}/a$. Dots denote derivatives with
respect to cosmic time and dashes correspond to conformal time
derivatives. Assuming $\mu^{2} \ll \xi R$, the solution of
(\ref{eomphi}) is:
\begin{equation}\label{phiproptoH}
\phi = \pm H \sqrt{ \frac{6}{\lambda}
\left[\epsilon(3-2\epsilon)-6\xi(2-\epsilon)\right]}\,.
\end{equation}
Clearly, $\phi \propto H$ and assumption (\ref{mHassumption}) is
satisfied. Moreover, observe we need $\xi <
\epsilon(3-2\epsilon)/\{6(2-\epsilon)\}$. The potential is
depicted in figure \ref{fig:potential}. Given a potential of the
form (\ref{potential}), we identify some interpretations leading
to this scenario. In the early universe the Higgs field satisfies
$\mu^{2} \ll |\xi R|$ and hence (\ref{phiproptoH}). A quintessence
field has $\mu^{2}=0$ and (\ref{phiproptoH}) is also satisfied in
the late universe. Finally, we can infer that a massless minimally
coupled scalar field is also allowed in the regime where
$0<\epsilon<3/2$ or $\epsilon>2$.

We work in the approximation where $\dot{\epsilon}=0$, which is
not equivalent to the standard slow-roll approximation of
inflation, where one introduces the slow-roll parameters
$\epsilon$ and $\eta$. These parameters are allowed to vary slowly
in time such that $\dot{\epsilon}$ and $\dot{\eta}$ are higher
order in the slow-roll parameters and can be neglected when one is
interested in the leading order behaviour only. When $\epsilon$ is
allowed to vary in time, equation (\ref{phiproptoH}) generalises
to $\phi=\alpha H$, where:
\begin{equation}\label{phiproptoH2}
\alpha = \pm \sqrt{ \frac{6}{\lambda}
\left[\epsilon(3-2\epsilon)+\frac{\dot{\epsilon}}{H}-6\xi(2-\epsilon)\right]}
+ \mathcal{O}\left(\dot{\alpha}\right) \,.
\end{equation}
The leading order slow-roll approximation to this equation
corresponds to:
\begin{equation}\label{phiproptoH3}
\alpha = \pm \sqrt{ \frac{6}{\lambda} \left[3 \epsilon
-6\xi(2-\epsilon)\right]} \,.
\end{equation}
Note the equation above depends on $\epsilon$, whereas it does not
depend on $\eta$.

\subsection{The Propagator}
\label{The Propagator}

Depending on which pole prescription one uses in the spirit of the
Schwinger-Keldysh formalism (see e.g. \cite{Prokopec:2003tm,
Weinberg:2005vy, Koksma:2007uq}), one can obtain different
propagators such as the anti-time ordered, time ordered or
Wightman propagators. We are primarily interested in the time
ordered or Feynman fermion propagator, defined by:
\begin{eqnarray}\label{Feynmanpropagator}
iS_{\mathrm{F}}^{ab} (x,\tilde{x}) &=& \langle \Omega |
\mathrm{T}\{ \hat{\psi}_{a}(x)\hat{\bar{\psi}}_{b}( \tilde{x})\} |
\Omega \rangle
\nonumber \\
&=& \theta(\eta - \tilde{\eta})
\langle \Omega |\hat{\psi}_{a}(x)\hat{\bar{\psi}}_{b}(\tilde{x})|\Omega\rangle \\
&& \qquad - \theta(\tilde{\eta} - \eta) \langle \Omega |
\hat{\bar{\psi}}_{b}(\tilde{x})\hat{\psi}_{a}(x)|\Omega\rangle
\nonumber \,.
\end{eqnarray}
The symbols $a$ and $b$ are the spinor indices and
$|\Omega\rangle$ denotes the state of the system. Also note the
minus sign in front of the second $\theta$-function. The fermion
propagator at tree level satisfies:
\begin{equation}\label{fermionprop}
\sqrt{-g} \left[ i \gamma^{\mu} \nabla_{\mu}^{x} - m \right]
iS_{\mathrm{F}}^{ab} (x,\tilde{x}) = i
\delta^{\scriptscriptstyle{D}}\!(x-\tilde{x}) \mathbb{I}^{ab}\,,
\end{equation}
and a likewise equation when $x$ and $\tilde{x}$ are interchanged.

\subsection{Properties of FLRW spacetimes}
\label{Properties of FLRW spacetimes}

We already discussed that FLRW spacetimes can be characterised by
the scale factor $a$ as a function of either cosmic or conformal
time. There are however a few more important properties of this
spacetime that are worth mentioning here.

An important relation that we will use throughout this manuscript
is:
\begin{equation}\label{epsilonconstant}
a(\eta) \eta = - \frac{1}{H(\eta)(1-\epsilon)} \,.
\end{equation}
This equation is equivalent to assumption (\ref{epsilon}), see
\cite{Janssen:2008dp, Janssen:2008dw}. Let us define some relevant
geometrical functions:
\begin{subequations}
\label{y}
\begin{eqnarray}
y_{++}(x,\tilde{x}) &=&
\frac{\Delta x_{++}^{2}(x,\tilde{x})}{ \eta \tilde{\eta}} \label{y++} \\
&=& \frac{1}{ \eta \tilde{\eta}} \Big(- \left(\left|\eta -
\tilde{\eta} \right| - i \varepsilon \right)^{2} + \| \vec{x} -
\vec{\tilde{x}}\|^{2} \Big) \nonumber  \\
y_{+-}(x,\tilde{x}) &=&
 \frac{1}{  \eta \tilde{\eta}} \Big(- \left( \phantom{|}\eta -
 \tilde{\eta}\phantom{|}
+ i \varepsilon \right)^{2} + \| \vec{x} - \vec{\tilde{x}}\|^{2}
\Big)  \label{y+-} \\
y_{-+}(x,\tilde{x}) &=& \frac{1}{ \eta \tilde{\eta}} \Big(- \left(
\phantom{|} \eta - \tilde{\eta}\phantom{|}
 - i \varepsilon \right)^{2} + \| \vec{x} - \vec{\tilde{x}}\|^{2}
\Big) \label{y-+} \\
y_{--}(x,\tilde{x}) &=& \frac{1}{ \eta \tilde{\eta}} \Big(-
\left(\left|\eta - \tilde{\eta} \right| + i \varepsilon
\right)^{2} + \| \vec{x} - \vec{\tilde{x}}\|^{2}
\Big),\phantom{1111} \label{y--}
\end{eqnarray}
\end{subequations}
which all vanish near the lightcone. Here, $\varepsilon > 0$ in
(\ref{y}) refers to the Feynman or time ordered pole
prescription\footnote{Note that the $\varepsilon$ of the Feynman
pole prescription is unrelated to the ``slow roll'' parameter
$\epsilon$ in FLRW spacetimes.}. In de Sitter spacetime, i.e.:
when $\epsilon = 0$, the de Sitter invariant function $y_{++}
(x,\tilde{x}) $ is related to the geodesic length $l
(x,\tilde{x})$ as $y_{++}|_{\varepsilon=0} = 4 \sin^{2}(H l/2)$.

\section{FLRW Fermion propagator: \newline The Massless Case}
\label{FLRW Fermion propagator: The Massless Case}

Massless fermions in FLRW spacetimes are not difficult to deal
with. By making use of equation (\ref{covariantderivative2}) the
propagator for massless (conformal) fermions should satisfy:
\begin{equation}\label{fermionpropmassless}
a^{-\frac{D+1}{2}} i \gamma^{b}\partial_{b}\left(a^{\frac{D-1}{2}}
i S_{c}(x,\tilde{x}) \right)= \frac{i}{a^{\scriptscriptstyle{D}}}
\delta^{\scriptscriptstyle{D}}\! (x-\tilde{x}) \,.
\end{equation}
Let us recall the following identity:
\begin{equation}\label{deltafunction}
\partial^{2} \frac{1}{ \Delta x^{\scriptscriptstyle{D-2}}_{++}(x,\tilde{x})} = \frac{ 4
\pi^{\scriptscriptstyle{D/2}}}{\Gamma\left(D/2-1\right)} i
\delta^{\scriptscriptstyle{D}}\!(x - \tilde{x}) \,.
\end{equation}
We can thus immediately infer the solution of equation
(\ref{fermionpropmassless}):
\begin{equation}\label{fermionpropmassless2}
i S_{c}(x,\tilde{x}) = \left(a \tilde{a}\right)^{-\frac{D-1}{2}}
\frac{\Gamma\left(D/2-1\right)}{4 \pi^{\scriptscriptstyle{D/2}}} i
\gamma^{b}\partial_{b} \frac{1}{ \Delta
x^{\scriptscriptstyle{D-2}}_{++} (x,\tilde{x})} .
\end{equation}
Here, $\tilde{a}=a(\tilde{\eta})$. This completes the calculation
for the propagator of massless fermions in FLRW spacetimes in $D$
dimensions. This propagator is valid in any FLRW spacetime and
assumption (\ref{epsilon}) can be relaxed. Because massless
fermions are conformal in any dimension, their propagator is much
easier to calculate than the massless scalar propagator
\cite{Janssen:2007ht, Janssen:2008dp, Janssen:2008dw,
Janssen:2008px}.

\section{FLRW Fermion propagator: \newline The Massive Case}
\label{FLRW Fermion propagator: The Massive Case}

A massive fermion is not conformal and hence its propagator in
general contains some complicated mass dependence\footnote{Note in
\cite{Birrell:1982ix} there is an erroneous statement regarding
massive fermion propagators in curved spacetimes. It is argued on
page 87 that a fermionic propagator in any spacetime can be
related to the scalar field propagator in that spacetime. This is
not correct. The (spinorial) structure that arises when a
covariant derivative acts on a spinor (\ref{covariantderivative2})
is much more complicated than for scalar fields.}. We firstly
solve for the fermionic mode functions. We generalise the approach
outlined in \cite{Cotaescu:2001cv, Garbrecht:2006jm} to
incorporate FLRW spacetimes with constant $\epsilon$. Using these
mode functions, we then return to position space to construct the
Feynman propagator.

\subsection{Fermionic Mode Functions}

Let us firstly define the rescaled fermionic spinor:
\begin{equation}\label{fermionrescaled}
\chi(x)= a^{\frac{D-1}{2}}(\eta) \psi(x) \,.
\end{equation}
Keeping an eye on equation (\ref{covariantderivative2}) one can
easily check that this factor is chosen such to conveniently
transform the covariant derivative into a partial derivative.
Equation of motion (\ref{fermioneom}) thus reads:
\begin{equation}\label{fermioneom2}
i \gamma^{b} \partial_{b} \chi(x) - a m \chi(x) = 0 \,.
\end{equation}
The canonical momentum associated with $\hat{\psi}(x)$ follows
from promoting the variation of (\ref{fermionaction}) with respect
to $\dot{\psi}(x)$ to an operator. According to the usual moves in
quantum field theory, we impose anti-commutation relations between
$\hat{\psi}(x)$ and its associated canonical momentum:
\begin{equation} \label{anticommutation}
\{\hat{\psi}_{a}(\mathbf{x},t), a^{\scriptscriptstyle{D-1}}(t)
\hat{\psi}^{\ast}_{b}(\mathbf{y},t) \} =
\delta^{\scriptscriptstyle{D-1}}\!(\mathbf{x}-\mathbf{y})
\delta_{ab} \,,
\end{equation}
with the other anti-commutators vanishing. By making use of the
rescaling in equation (\ref{fermionrescaled}), appreciate that the
anti-commutation relations above simplify:
\begin{subequations}
\label{anticommutation2}
\begin{eqnarray}
\{\hat{\chi}_{a}(\mathbf{x},t),
\hat{\chi}^{\ast}_{b}(\mathbf{y},t) \} &=&
\delta^{\scriptscriptstyle{D-1}}\!(\mathbf{x}-\mathbf{y})
\delta_{ab}
\label{anticommutation2a} \\
\{\hat{\chi}_{a}(\mathbf{x},t), \hat{\chi}_{b}(\mathbf{y},t) \}
&=& 0
\label{anticommutation2b} \\
\{\hat{\chi}^{\ast}_{a}(\mathbf{x},t),
\hat{\chi}^{\ast}_{b}(\mathbf{y},t) \} &=& 0
\label{anticommutation2c} \,.
\end{eqnarray}
\end{subequations}

\subsubsection{Chirality and Helicity Decomposition}
\label{Chirality and Helicity Decomposition}

Therefore we expand the rescaled spinors $\hat{\chi}(x)$ and
$\hat{\bar{\chi}}(x)$ in creation and annihilation operators as
follows:
\begin{subequations}
\label{creationannihilationexpansion}
\begin{eqnarray}
\hat{\chi}(x) &=& \int
\frac{\mathrm{d}^{\scriptscriptstyle{D-1}}\mathbf{k}}{(2\pi)^{\scriptscriptstyle{D-1}}}
\sum_{h} \hat{a}_{\mathbf{k},h}\chi^{(h)}(\mathbf{k},\eta) e^{i
\mathbf{k}\cdot \mathbf{x}} \label{creationannihilationexpansiona}
\\
&& \qquad\qquad\qquad + \phantom{1} \hat{b}_{\mathbf{k},h}^{\dag}
\nu^{(h)}(\mathbf{k},\eta) e^{-i
\mathbf{k}\cdot \mathbf{x}} \nonumber \\
\hat{\bar{\chi}}(x) &=& \int
\frac{\mathrm{d}^{\scriptscriptstyle{D-1}}\mathbf{k}}{(2\pi)^{\scriptscriptstyle{D-1}}}
\sum_{h} \hat{b}_{\mathbf{k},h}\bar{\nu}^{(h)}(\mathbf{k},\eta)
e^{i \mathbf{k}\cdot \mathbf{x}}
\label{creationannihilationexpansionb}
\\
&& \qquad\qquad\qquad + \phantom{1} \hat{a}_{\mathbf{k},h}^{\dag}
\bar{\chi}^{(h)}(\mathbf{k},\eta) e^{-i \mathbf{k}\cdot
\mathbf{x}} \nonumber \,.
\end{eqnarray}
\end{subequations}
The expansion above merits a few remarks. Firstly,
$\hat{a}_{\mathbf{k},h}$ and $\hat{b}_{\mathbf{k},h}$ are the
fermion and anti-fermion annihilation operators of helicity $h$,
respectively, in the usual sense:
$\hat{a}_{\mathbf{k},h}|\Omega\rangle = 0 =
\hat{b}_{\mathbf{k},h}|\Omega\rangle$. The helicity $h$, i.e.: the
spin in the direction of motion, can be either $+1$ or $-1$ in
units of $\hbar$.

Let us for simplicity return to the $D=4$ setting we are familiar
with. We will generalise the following considerations shortly to
arbitrary dimensions.

In the equation above $\chi^{(h)}(\mathbf{k},\eta)$ is a 4-spinor
of helicity $h$. We decompose the 4-spinor $\chi(\mathbf{k},\eta)$
into a direct product of chirality and helicity 2-spinors:
\begin{equation}\label{fermiondecompose}
\chi(\mathbf{k},\eta)= \sum_{h} \chi^{(h)}(\mathbf{k},\eta) =
\sum_{h}
\begin{pmatrix}
\chi_{\mathrm{\scriptscriptstyle{L}},h}(\mathbf{k},\eta) \\
\chi_{\mathrm{\scriptscriptstyle{R}},h}(\mathbf{k},\eta)
\end{pmatrix}
\otimes \xi_{h} \,.
\end{equation}
Here, $\chi_{\mathrm{\scriptscriptstyle{L}},h}(\mathbf{k},\eta)$
and $\chi_{\mathrm{\scriptscriptstyle{R}},h}(\mathbf{k},\eta)$ are
left- and right-handed 1-spinors of helicity $h$, respectively.
Furthermore, $\xi_{h}$ is the helicity 2-eigenspinor:
\begin{equation}\label{helicity}
\hat{h} \xi_{h} \equiv \hat{\vec{k}} \cdot \vec{\sigma} \xi_{h} =
h \xi_{h} \,,
\end{equation}
where $\vec{\sigma} = (\sigma^1,\sigma^2,\sigma^3)$ is a shorthand
notation for the Pauli matrices. Furthermore note that $\hat{h}^2=
\hat{h}^{\dag}\hat{h} =1$.

For future convenience, let us explicitly derive the two vectors
$\xi_{h}$. We write:
\begin{equation}
\label{kvector1} \hat{\vec{k}} =
(\hat{k}_{x},\hat{k}_{y},\hat{k}_{z}) \,,
\end{equation}
and keep in mind that:
\begin{equation}
\label{kvector2} \| \hat{\vec{k}} \|^{2} = \hat{k}_{x}^{2} +
\hat{k}_{y}^{2} + \hat{k}_{z}^{2} =1 \,.
\end{equation}
It is a trivial exercise to solve for the two eigenvectors of
$\hat{h}$, which correctly normalised to unity read:
\begin{subequations}
\label{helicity1}
\begin{eqnarray}
\xi_{+} &=& \frac{1}{\sqrt{2(1-\hat{k}_{z})}}
\begin{pmatrix}
\hat{k}_{x} - i \hat{k}_{y}\\
1-\hat{k}_{z}
\end{pmatrix} \label{helicity1a} \\
\xi_{-} &=& \frac{1}{\sqrt{2(1+\hat{k}_{z})}}
\begin{pmatrix}
i \hat{k}_{y} -  \hat{k}_{x}\\
1+\hat{k}_{z}
\end{pmatrix} \label{helicity1b}
\,.
\end{eqnarray}
\end{subequations}
Moreover, the helicity eigenstates are mutually orthogonal:
\begin{equation} \label{kvectorsorthogonal}
\xi_{+}^{\dag} \cdot \xi_{-} = 0 \,.
\end{equation}
Finally, we have tacitly ignored the expansion of the
anti-particle contribution $\nu(\mathbf{k},\eta)$ in chirality and
helicity spinors in equation
(\ref{creationannihilationexpansion}). We will return to this
subtlety shortly.

\subsubsection{Generalisation to Higher Dimensions}
\label{Generalisation to Higher Dimensions}

Our chirality and helicity decomposition of the spinor degrees of
freedom in equation (\ref{fermiondecompose}) also works in $D$
spacetime dimensions. Once a particular representation of the
gamma matrices has been found, the projection operator:
\begin{equation}\label{P+-}
P_{\pm} = \frac{1 \pm \gamma^{\scriptscriptstyle{D}+1}}{2}\,,
\end{equation}
splits the $2^{\scriptscriptstyle{D}/2}$-spinor degrees of freedom
in two equal contributions of definite chirality, i.e.: a left-
and right-handed $2^{(\scriptscriptstyle{D-2})/2}$-spinor (see
e.g. \cite{DeWit:1986it}). Here:
\begin{equation}\label{P+-2}
\gamma^{\scriptscriptstyle{D}+1} = \alpha_{\scriptscriptstyle{D}}
\gamma^{0}\cdots\gamma^{\scriptscriptstyle{D-1}}\,,
\end{equation}
where $\alpha_{\scriptscriptstyle{D}}$ is fixed by requiring that
$P_{\pm}$ is a proper projection operator:
\begin{equation}\label{P+-3}
P_{\pm}^{2}=P_{\pm} \,\,\, , \,\,\, P_{+}P_{-}=0 \quad
\Longrightarrow \quad (\gamma^{\scriptscriptstyle{D}+1})^{2} = 1
\,.
\end{equation}
This yields:
\begin{equation}\label{P+-4}
\alpha_{\scriptscriptstyle{D}} = \exp\left[ i
\frac{\pi}{4}(D-1)(D+2)\right] \,.
\end{equation}
In $D=4$, we recognise the familiar $\gamma^{5}= i
\gamma^{0}\gamma^{1}\gamma^{2}\gamma^{3}$. An helicity operator
that commutes with the projection operator can be defined in
Fourier space as:
\begin{equation}\label{helicityD}
\hat{h} = -
\frac{e^{i\pi\frac{(D-2)(D-1)}{4}}}{\Gamma(D-1)}\epsilon^{i l_{1}
\cdots l_{\scriptscriptstyle{D-2}}} k_{i} \gamma^{l_{1}} \cdots
\gamma^{l_{\scriptscriptstyle{D-2}}}\,,
\end{equation}
where $\epsilon^{i l_{1} \cdots l_{\scriptscriptstyle{D-2}}}$ is
the Levi-Civita tensor in $D-1$ dimensions. The exponential phase
ensures that $\hat{h}$ is hermitian. Finally, the Gamma function
has been inserted to account for the number of different
permutations of gamma matrices.

Hence, $\chi_{\mathrm{\scriptscriptstyle{L}}, h}
(\mathbf{k},\eta)$ and $\chi_{\mathrm{\scriptscriptstyle{R}}, h}
(\mathbf{k},\eta)$ are
\mbox{$2^{(\scriptscriptstyle{D-4})/2}$-spinors}. Note the
definition of the helicity operator is consistent with
(\ref{helicity}). Most importantly, the chirality and helicity
decomposition as proposed in equation (\ref{fermiondecompose})
carries through for any dimension.

When an explicit form of $\xi_{h}$ is required, we return to $D=4$
and use equation (\ref{helicity1}). We subsequently generalise the
result of such calculations by analytically extending it to
arbitrary $D$.

\subsubsection{Spinorial Normalisation Conditions}
\label{Spinorial Normalisation Conditions}

In textbooks on quantum field theory, such as
\cite{Peskin:1995ev,Itzykson:1980rh}, one expands the fermion in
momentum space in spin eigenstates, whereas we expand in helicity
eigenstates. This results in a different normalisation requirement
on the spinors in momentum space and consequently the standard
textbook results cannot be straightforwardly copied. The reason
for expanding in helicity eigenstates is that in curved spacetimes
helicity is more convenient to work with rather than spin.

Spin is a conserved quantity, i.e.: an appropriate quantum number,
in static backgrounds, whereas helicity is conserved in time
dependent but spatially homogeneous backgrounds
\cite{Prokopec:2003pj}. One can easily show that the helicity
operator constructed above in (\ref{helicityD}) commutes with the
kinetic operator in Wigner space for spatially homogeneous
correlators.

We normalise the spinors using two conditions: a consistent
canonical quantisation and orthogonality of particle and
anti-particle spinors.

We thus impose the usual anti-commutation relations between
creation and annihilation operators:
\begin{subequations}
\label{anticommutation3}
\begin{eqnarray}
\{ \hat{a}_{\mathbf{k},h}, \hat{a}_{\mathbf{k}',h'}^{\dag} \} &=&
(2\pi)^{\scriptscriptstyle{D-1}}
\delta^{\scriptscriptstyle{D-1}}(\mathbf{k}-\mathbf{k}')
\delta_{h,h'}
\label{anticommutation3a} \\
\{ \hat{b}_{\mathbf{k},h}, \hat{b}_{\mathbf{k}',h'}^{\dag} \} &=&
(2\pi)^{\scriptscriptstyle{D-1}}
\delta^{\scriptscriptstyle{D-1}}(\mathbf{k}-\mathbf{k}')
\delta_{h,h'} \label{anticommutation3b} \,,
\end{eqnarray}
\end{subequations}
and all other anti-commutators vanish identically. Now
(\ref{anticommutation3}) is consistent with
(\ref{anticommutation2}) if and only if:
\begin{equation}
\label{consistentquantisation} \sum_{h}
\chi^{(h)}_{a}(\mathbf{k},\eta) \chi^{\ast
(h)}_{b}(\mathbf{k},\eta) + \nu^{(h)}_{a}(-\mathbf{k},\eta)
\nu^{\ast (h)}_{b}(-\mathbf{k},\eta) = \delta_{ab} .
\end{equation}
Note that this is a condition on $2^{(D/2)-1}(2^{D/2}+1)$ matrix
elements in spinor space, i.e.: for each $a$ and $b$ the above
equality has to be satisfied.

Secondly, we require that the particle and anti-particle spinors
are mutually orthogonal:
\begin{equation} \label{orthogonality}
\bar{\chi}^{(h)}(\mathbf{k},\eta) \nu^{(h')}(\mathbf{k},\eta) = 0
= \bar{\nu}^{(h')}(\mathbf{k},\eta) \chi^{(h)}(\mathbf{k},\eta)
\,.
\end{equation}
This restricts 4 matrix elements in helicity space, i.e.: for each
$h,h'= \pm$ the above equality needs to be obeyed.

Peskin and Schroeder \cite{Peskin:1995ev} impose normalisation
conditions on both particle and anti-particle spinors in spin
space. The corresponding conditions in helicity space are only
partially satisfied. Both particle and anti-particle spinors of
different helicity are trivially orthogonal by construction
(\ref{kvectorsorthogonal}). However, we do not relate the
amplitude of equal helicity states for particles and
anti-particles to the mass of the spinor because helicity
eigenstates for a particle at rest cannot be constructed.

\subsubsection{Particle Mode Functions}
\label{Particle Mode Functions}

We will now solve the Dirac equation. Let us insert equation
(\ref{creationannihilationexpansion}) into the Dirac equation
(\ref{fermioneom2}). Making use of our chirality and helicity
decomposition (\ref{fermiondecompose}) transforms the Dirac
equation to a first order coupled system of differential
equations:
\begin{subequations}
\label{DiracEq1}
\begin{eqnarray}
i \chi'_{\mathrm{\scriptscriptstyle{L}},h} + h k
\chi_{\mathrm{\scriptscriptstyle{L}},h} - a m
\chi_{\mathrm{\scriptscriptstyle{R}},h} &=& 0
\label{DiracEq1a} \quad \\
i \chi'_{\mathrm{\scriptscriptstyle{R}},h} - h k
\chi_{\mathrm{\scriptscriptstyle{R}},h} - a m
\chi_{\mathrm{\scriptscriptstyle{L}},h} &=& 0 \label{DiracEq1b}
\,. \quad
\end{eqnarray}
\end{subequations}
The arguments of the functions in the equation above have been
omitted for notational convenience. The left- and right-handed
spinors however are now a function of the magnitude of the Fourier
mode $k$ only, because the differential equation is invariant
under $\vec{k} \rightarrow - \vec{k}$.

Standard model fermions are chiral, in the sense that
$m_{\mathrm{\scriptscriptstyle{R}}} \neq
m_{\mathrm{\scriptscriptstyle{L}}}$. For simplicity, we take one
mass $m = m_{\mathrm{\scriptscriptstyle{R}}} =
m_{\mathrm{\scriptscriptstyle{L}}}$ only. The following derivation
can easily be generalised when standard model fermions are
considered by making use of the $P_{\pm}$ projectors in equation
(\ref{P+-}).

We now return to the expansion in chirality and helicity spinors
of the anti-particle contribution $\nu(\mathbf{k},\eta)$, which we
did not need so far. Since the Dirac equation is a linear
differential equation, the solution for $\nu(\mathbf{k},\eta)$
cannot contain any new degrees of freedom once we completely
solved for $\chi(\mathbf{k},\eta)$. In other words: the form of
$\nu(\mathbf{k},\eta)$ is dictated once we have completely
specified $\chi(\mathbf{k},\eta)$. We expand
$\nu(\mathbf{k},\eta)$ slightly differently:
\begin{equation}\label{fermiondecompose2}
\nu(\mathbf{k},\eta)= \sum_{h} \nu^{(h)}(\mathbf{k},\eta) =
\sum_{h}
\begin{pmatrix}
\nu_{\mathrm{\scriptscriptstyle{R}},h}(\mathbf{k},\eta) \\
\nu_{\mathrm{\scriptscriptstyle{L}},h}(\mathbf{k},\eta)
\end{pmatrix}
\otimes \xi_{h} \,.
\end{equation}
Note we flipped the position of the left- and right-handed spinors
compared to (\ref{fermiondecompose}). Consequently, the resulting
equations of motion for both
$\nu_{\mathrm{\scriptscriptstyle{R}},h}(\mathbf{k},\eta)$ and
$\nu_{\mathrm{\scriptscriptstyle{L}},h}(\mathbf{k},\eta)$ are
identical to (\ref{DiracEq1}).

Alternatively, we could have expanded differently and flipped
helicity, i.e.: send $h\rightarrow -h$ in equation
(\ref{fermiondecompose}). Because the mass enforces mixing between
different chirality states, labelling these states differently has
no physical relevance.

We return to equation of motion (\ref{DiracEq1}) and define:
\begin{equation}\label{uplusminus}
u_{\pm h}(k,\eta) \equiv
\frac{\chi_{\mathrm{\scriptscriptstyle{L}},h}(k,\eta) \pm
\chi_{\mathrm{\scriptscriptstyle{R}},h}(k,\eta)}{\sqrt{2}}  \,.
\end{equation}
This transforms equation (\ref{DiracEq1}) to:
\begin{subequations}
\label{DiracEq2}
\begin{eqnarray}
i u'_{+ h} + h k u_{- h} - a m u_{+ h} &=& 0
\label{DiracEq2a} \quad \\
i u'_{- h} + h k u_{+ h} + a m u_{- h} &=& 0 \label{DiracEq2b} \,.
\quad
\end{eqnarray}
\end{subequations}
We exploit assumption (\ref{epsilon}), which is equivalent to
(\ref{epsilonconstant}). Inserting this expression into the
equation above yields:
\begin{subequations}
\label{DiracEq3}
\begin{eqnarray}
i u'_{+ h} + h k u_{- h}+ \frac{m}{H(1-\epsilon)} \frac{u_{+
h}}{\eta} &=& 0
\label{DiracEq3a}  \\
i u'_{- h}+ h k u_{+ h} - \frac{m}{H(1-\epsilon)} \frac{u_{-
h}}{\eta} &=& 0\,. \label{DiracEq3b}
\end{eqnarray}
\end{subequations}
Recall that the mass of a fermion in Yukawa theory is time
dependent in FLRW spacetimes. There does not exist a rescaling of
the functions $u_{\pm h}$ that removes the mass dependence in the
equation of motion above. Even when one assumes that the mass $m$
is time independent, one cannot find such a rescaling. It is
particularly simple to choose $m/H$ constant as following from
equation (\ref{phiproptoH}) and for convenience we define:
\begin{equation}\label{zeta}
\zeta = \frac{m}{H(1-\epsilon)} \,.
\end{equation}
We can proceed and decouple the two linear differential equations
to find:
\begin{equation}\label{DiracEq4}
u_{\pm h}'' + \left(k^{2} +  \frac{\zeta^2 \pm i
\zeta}{\eta^2}\right) u_{\pm h} = 0 \,.
\end{equation}
The solutions are given by:
\begin{equation}\label{DiracEq5sols}
u_{\pm h} = \alpha_{\pm k}^{h} \sqrt{-k\eta} \,
H^{(1)}_{\nu_{\pm}}(-k\eta) \,,
\end{equation}
where $\alpha_{\pm k}^{h}$ are two normalisation constants that
still need to be determined and $H^{(1)}_{\nu_{\pm}}$ is the
Hankel function of the first kind of order:
\begin{equation}\label{order}
\nu_{\pm} = \frac{1}{2} \mp i \zeta \,.
\end{equation}
This implies:
\begin{subequations}
\begin{eqnarray}
\nu_{+}+\nu_{-} &=& 1 \label{Hankelprops6}\\
\nu_{+}^{\ast} &=& \nu_{-} \label{Hankelprops7}\,.
\end{eqnarray}
\end{subequations}
The Hankel function of the second kind also solves equation
(\ref{DiracEq4}). In the infinite asymptotic past, equivalent to
the deep UV, the fermions become effectively massless for
$\epsilon<1$ and the distinction between $u_{+ h}$ and $u_{- h}$
vanishes as a consequence. Therefore we require that the vacuum
mode functions in this regime equal the standard conformal vacuum
solutions:
\begin{subequations}
\label{UVlimit}
\begin{eqnarray}
\lim_{\eta \rightarrow - \infty} u_{+ h}(-k\eta) &=&
\frac{1}{\sqrt{2}} \, e^{-i k \eta} \label{UVlimita}\\
\lim_{\eta \rightarrow - \infty} u_{- h}(-k\eta) &=& \frac{- h
}{\sqrt{2}} \, e^{-i k \eta} \label{UVlimitb}\,.
\end{eqnarray}
\end{subequations}
This excludes the Hankel function of the second kind for the
moment from contributing. Away from the vacuum, we should of
course also incorporate the second solution to (\ref{DiracEq4}) to
allow for mode mixing. Note that the UV limits above yield the
familiar UV behaviour for the left- and right-handed spinors
(e.g.: \cite{Peskin:1995ev}):
\begin{subequations}
\label{UVlimit2}
\begin{eqnarray}
\lim_{\eta \rightarrow - \infty}
\chi_{\mathrm{\scriptscriptstyle{L}},h}(k,\eta) &=&
\frac{1-h}{2} \, e^{-i k \eta} \label{UVlimit2a}\\
\lim_{\eta \rightarrow - \infty}
\chi_{\mathrm{\scriptscriptstyle{R}},h}(k,\eta) &=& \frac{1+ h}{2}
\, e^{-i k \eta} \label{UVlimit2b}\,.
\end{eqnarray}
\end{subequations}
When solutions (\ref{DiracEq5sols}) are substituted in
(\ref{DiracEq3a}) one finds:
\begin{equation}\label{normalisation1}
\alpha_{- k}^{h} = i h \alpha_{+ k}^{h} \, e^{i\pi\nu_{-}} \,.
\end{equation}
We are free to fix the remaining coefficient $\alpha_{+ k}^{h}$ by
the following condition:
\begin{equation}\label{normalisation2}
\sum_{\pm} \left| u_{\pm h} \right|^2 = \left| u_{+ h} \right|^2 +
\left| u_{- h} \right|^2 = 1 \,.
\end{equation}
We postpone the motivation for this requirement to section
\ref{Normalisation of the Mode Functions}. For the moment, let us
simply normalise accordingly for convenience. We thus find:
\begin{equation}\label{normalisation3}
 \alpha_{+ k}^{h} =  \sqrt{\frac{\pi}{4}} \, e^{i
\frac{\pi}{2} ( \nu_{+} + 1/2)} \,,
\end{equation}
where we have chosen the unobservable phase factor in accordance
with the UV limit (\ref{UVlimit}). We use some well-known
identities involving Hankel functions which we list for
convenience in appendix \ref{Properties of Hankel Functions}.

Concluding, we find that the vacuum solutions to equation
(\ref{DiracEq3}) neglecting derivatives of $\epsilon$ and treating
the quotient $m/H$ as time independent are given by:
\begin{subequations}
\label{DiracEq6sols}
\begin{eqnarray}
u_{+ h}(-k\eta) &=& \phantom{-h} e^{i\frac{\pi}{2}(\nu_{+}+1/2)}
\,\sqrt{-\frac{\pi k \eta}{4}} \, H^{(1)}_{\nu_{+}}(-k\eta)
\phantom{aaa}
\label{DiracEq6asols}  \\
u_{- h}(-k\eta) &=& -h e^{i\frac{\pi}{2}(\nu_{-}+1/2)}
\,\sqrt{-\frac{\pi k \eta}{4}} \, H^{(1)}_{\nu_{-}}(-k\eta) \,.
\phantom{aaa} \label{DiracEq6bsols}
\end{eqnarray}
\end{subequations}
As we send $\epsilon \rightarrow 0$ and approach de Sitter space,
our solution is in accordance with \cite{Garbrecht:2006jm}, as can
be seen from (\ref{order}). In this solution we have tacitly
assumed that $\epsilon < 1$ such that $\eta<0$, equivalent to an
accelerated expansion. If however $\epsilon
> 1$ or $\eta>0$, which corresponds to a decelerating universe, one should
simply use relations (\ref{Hankelprops8}--\ref{Hankelprops9}) in
appendix \ref{Properties of Hankel Functions}. The case $\epsilon
= 1$ requires special attention, see e.g. \cite{Janssen:2008px}
for the analogous scalar propagator case.

Let us now turn our attention to finding the solution for the
second mode function corresponding to the excited state. The
procedure is completely analogous: we again let $\eta \rightarrow
- \infty$ and match to the appropriate UV expansion involving the
second solution $H_{\nu_{\pm}}^{(2)}(-k\eta)$. We normalise the
solution according to (\ref{normalisation2}) and choose the phase
factor in accordance with the corresponding UV limit. We can thus
construct the total solution by linear superposition:
\begin{subequations}
\label{DiracEqtotalsol1}
\begin{eqnarray}
u_{+ h}(-k\eta) &=& c \mathcal{H}(\eta,\zeta) + d \tilde{\mathcal{H}}(\eta,\zeta) \label{DiracEqtotalsol1a}  \\
u_{- h}(-k\eta) &=& -h c \tilde{\mathcal{H}}^{\ast} (\eta,\zeta) +
h d \mathcal{H}^{\ast}(\eta,\zeta)\,, \label{DiracEqtotalsol1b}
\end{eqnarray}
\end{subequations}
where:
\begin{subequations}
\label{DiracEqtotalsol2}
\begin{eqnarray}
\mathcal{H}(\eta,\zeta) &=& e^{i\frac{\pi}{2}(\nu_{+}+1/2)}
\,\sqrt{-\frac{\pi k \eta}{4}} \, H^{(1)}_{\nu_{+}}(-k\eta)
\label{DiracEqtotalsol2a}  \\
\tilde{\mathcal{H}}(\eta,\zeta) &=&
e^{-i\frac{\pi}{2}(\nu_{+}+1/2)} \,\sqrt{-\frac{\pi k \eta}{4}} \,
H^{(2)}_{\nu_{+}}(-k\eta) \,. \label{DiracEqtotalsol2b}
\end{eqnarray}
\end{subequations}
Note that we have omitted all $k$ dependence in
(\ref{DiracEqtotalsol1}) for notational convenience. Note finally
that the two fundamental solutions (\ref{DiracEqtotalsol2}) in
$u_{+ h}(-k\eta)$ are not complex conjugates of each other, unlike
for example for the scalar field case. However, the second
solution $u_{- h}(-k\eta)$ can be obtained from complex
conjugation of the fundamental solutions comprising $u_{+
h}(-k\eta)$, as can clearly be seen in equation
(\ref{DiracEqtotalsol1}) above.

For the moment, we leave the normalisation constants $c$ and $d$
undetermined. Appreciate we are nevertheless free to normalise the
fundamental solutions to unity. Of course we will determine the
normalisation constants shortly by applying our analysis performed
in subsection \ref{Spinorial Normalisation Conditions}. But first,
we need to solve for the anti-particle mode functions.

\subsubsection{Anti-particle Mode Functions}
\label{Anti-particle Mode Functions}

Let us stress again that the other spinor
$\nu^{(h)}(\mathbf{k},\eta)$ cannot contain any new degrees of
freedom because Dirac's equation is linear. We only have to
determine the connection with $\chi^{(h)}(\mathbf{k},\eta)$. When
we expand according to (\ref{fermiondecompose2}) the coupled
linear differential equations
$\nu_{\mathrm{\scriptscriptstyle{R}},h}(\mathbf{k},\eta)$ and
$\nu_{\mathrm{\scriptscriptstyle{L}},h}(\mathbf{k},\eta)$ obey are
identical to (\ref{DiracEq1}). Therefore, we can analogously
define:
\begin{equation}\label{nuplusminus}
\nu_{\pm h}(k,\eta) \equiv
\frac{\nu_{\mathrm{\scriptscriptstyle{L}},h}(k,\eta) \pm
\nu_{\mathrm{\scriptscriptstyle{R}},h}(k,\eta)}{\sqrt{2}}  \,.
\end{equation}
We are rewarded for our extensive discussion of chirality and
helicity. The solutions of $\nu_{\pm h}(-k\eta)$ are identical to
(\ref{DiracEqtotalsol1}), where we should only denote the
coefficients in this solution differently:
\begin{subequations}
\label{DiracEqtotalsol5}
\begin{eqnarray}
\nu_{+ h}(-k\eta) &=& f \mathcal{H}(\eta,\zeta) + g
\tilde{\mathcal{H}}(\eta,\zeta) \label{DiracEqtotalsol5a}  \\
\nu_{- h}(-k\eta) &=& -h f \tilde{\mathcal{H}}^{\ast} (\eta,\zeta)
+ h g \mathcal{H}^{\ast}(\eta,\zeta)\,. \label{DiracEqtotalsol5b}
\end{eqnarray}
\end{subequations}
The functions appearing in this equation are given by
(\ref{DiracEqtotalsol2}). We need to determine how the
coefficients $c$ and $d$ of $u_{\pm h}(-k\eta)$ relate to $f$ and
$g$ of $\nu_{\pm h}(-k\eta)$. Moreover, we need to find the
correct normalisation condition for $c$ and $d$.

\subsubsection{Normalisation of the Mode Functions}
\label{Normalisation of the Mode Functions}

Inspired by our discussion in \ref{Spinorial Normalisation
Conditions}, we impose two normalisation conditions on the
spinors: a consistent canonical quantisation
(\ref{consistentquantisation}) and orthogonality between particle
and anti-particle spinors (\ref{orthogonality}).

After quite some work the, naively, 10 constraint equations in
$D=4$ in (\ref{consistentquantisation}) yield the following 3
conditions on our coefficients:
\begin{subequations}
\label{consistentquantisation2}
\begin{eqnarray}
|c|^{2}+ |d|^{2}+ |f|^{2}+ |g|^{2} &=& 2
\label{consistentquantisation2a} \\
|c|^{2}- |d|^{2}+ |f|^{2}- |g|^{2} &=& 0
\label{consistentquantisation2b} \\
c d^{\ast} + f g^{\ast} &=& 0 \label{consistentquantisation2c} \,.
\end{eqnarray}
The orthogonality condition (\ref{orthogonality}) yields one more
condition:
\begin{eqnarray}
c f^{\ast} + d g^{\ast} &=& 0 \label{consistentquantisation2d} \,.
\end{eqnarray}
\end{subequations}
The solution is:
\begin{subequations}
\label{consistentquantisation3}
\begin{eqnarray}
|f| &=& |d|
\label{consistentquantisation3a} \\
|g| &=& |c|
\label{consistentquantisation3b} \\
\phi_{f} &=& \phi_{c} - \phi_{d} + \phi_{g} \pm \pi
\label{consistentquantisation3c} \,.
\end{eqnarray}
The last line relates the phases of our normalisation constants,
e.g.: $\phi_{f}$ denotes the phase of $f$. Moreover, from
(\ref{consistentquantisation2}) we can derive:
\begin{eqnarray}
|c|^{2} + |d|^{2} &=& 1 \label{consistentquantisation3d} \\
|f|^{2} + |g|^{2} &=& 1 \label{consistentquantisation3e} \,,
\end{eqnarray}
\end{subequations}
where the second line is a consequence of the first. Note that
this condition also follows from charge conservation as argued in
\cite{Garbrecht:2006jm}. Changing the normalisation in
(\ref{normalisation2}) results in a change of the right-hand side
of equations (\ref{consistentquantisation2a}),
(\ref{consistentquantisation3d}) and
(\ref{consistentquantisation3e}). Normalising the fundamental
solutions to unity as in (\ref{normalisation2}) is particularly
convenient as it allows for a particle interpretation of $|d|^{2}$
and $|f|^{2}$.

The normalisation constants provide us with three physical degrees
of freedom. Of course $|c|$ is physical which determines $|d|$
through (\ref{consistentquantisation3d}). Hence, $\phi_{d}$ is
physical just as in the scalar field case. Since $\phi_{c}$ is a
phase that can be removed without physical consequences, the phase
relation (\ref{consistentquantisation3c}) determines $\phi_{f}$ in
terms of the third physical phase $\phi_{g}$. Note finally that
one of the two phase relations (\ref{consistentquantisation2c})
and (\ref{consistentquantisation2d}) is redundant in this part of
the analysis.

A final remark is in order. To derive
(\ref{consistentquantisation2}) we needed the explicit form of
$\xi_{h}$ in equation (\ref{helicity1}), which is only valid in
$D=4$. However, we analytically continue this result to arbitrary
dimension and one naturally finds (\ref{consistentquantisation2}).

\subsubsection{Summary}
\label{Summary}

Using the solutions for $u_{\pm h}(k,\eta)$ and $\nu_{\pm
h}(k,\eta)$ given in (\ref{DiracEqtotalsol1}) and
(\ref{DiracEqtotalsol5}) respectively, we recall relations
(\ref{uplusminus}) and (\ref{nuplusminus}) to find:
\begin{subequations}
\label{summary1}
\begin{eqnarray}
\chi_{\mathrm{\scriptscriptstyle{L}},h}(k,\eta) &=&
\frac{1}{\sqrt{2}}\left[ c\{ \mathcal{H} - h
\tilde{\mathcal{H}}^{\ast} \} + d\{ \tilde{\mathcal{H}} + h
\mathcal{H}^{\ast} \}\right] \phantom{1111}
\label{summary1a} \\
\chi_{\mathrm{\scriptscriptstyle{R}},h}(k,\eta) &=&
\frac{1}{\sqrt{2}}\left[ c\{ \mathcal{H} + h
\tilde{\mathcal{H}}^{\ast} \} + d\{ \tilde{\mathcal{H}} - h
\mathcal{H}^{\ast} \} \right] \label{summary1b} , \phantom{1111}
\end{eqnarray}
and moreover:
\begin{eqnarray}
\nu_{\mathrm{\scriptscriptstyle{L}},h}(k,\eta) &=&
\frac{1}{\sqrt{2}}\left[ f\{ \mathcal{H} - h
\tilde{\mathcal{H}}^{\ast} \} + g\{ \tilde{\mathcal{H}} + h
\mathcal{H}^{\ast}\}\right]\phantom{1111}
\label{summary1c} \\
\nu_{\mathrm{\scriptscriptstyle{R}},h}(k,\eta) &=&
\frac{1}{\sqrt{2}}\left[ f\{ \mathcal{H} + h
\tilde{\mathcal{H}}^{\ast} \} + g\{ \tilde{\mathcal{H}} - h
\mathcal{H}^{\ast} \}\right] , \phantom{1111}\label{summary1d}
\end{eqnarray}
\end{subequations}
where the fundamental solution are given in equation
(\ref{DiracEqtotalsol2}). The normalisation conditions
(\ref{consistentquantisation3}) determine $f$ and $g$ in terms of
$c$ and $d$. Moreover, we have $|c|^{2}+ |d|^{2}=1$.

\subsection{The Feynman Propagator}
\label{The Feynman Propagator}

Having discussed the fermionic mode functions in some detail, we
can turn our attention to solving the Feynman propagator for
fermions. We insert the rescaled spinors, expanded in terms of
creation and annihilation operators, into the formal definition of
the Feynman propagator (\ref{Feynmanpropagator}), which yields:
\begin{eqnarray}\label{Feynmanpropagator2}
&& iS_{\mathrm{F}}^{ab} (x,\tilde{x}) = \\
&& \qquad \theta (\eta-\tilde{\eta}) a^{-\frac{D-1}{2}}
\tilde{a}^{-\frac{D-1}{2}} \nonumber \\
&& \qquad\quad \times \int
\frac{\mathrm{d}^{\scriptscriptstyle{D-1}}\mathbf{k}}{(2\pi)^{\scriptscriptstyle{D-1}}}
\sum_{h} \chi_{a}^{(h)}(\mathbf{k}, \eta)
\bar{\chi}_{b}^{(h)}(\mathbf{k}, \tilde{\eta}) e^{ i
\mathbf{k}\cdot(\mathbf{x}-\tilde{\mathbf{x}})} \nonumber \\
&& \quad - \theta (\tilde{\eta} - \eta) a^{-\frac{D-1}{2}}
\tilde{a}^{-\frac{D-1}{2}} \nonumber \\
&& \qquad\quad \times \int
\frac{\mathrm{d}^{\scriptscriptstyle{D-1}}\mathbf{k}}{(2\pi)^{\scriptscriptstyle{D-1}}}
\sum_{h}
 \bar{\nu}_{b}^{(h)}(\mathbf{k}, \tilde{\eta}) \nu_{a}^{(h)}(\mathbf{k},\eta)  e^{ - i
\mathbf{k}\cdot(\mathbf{x}-\tilde{\mathbf{x}})}, \nonumber
\end{eqnarray}
where $a=a(\eta)$ and $\tilde{a}=a(\tilde{\eta})$. The reader can
easily see that we have to make use of the solutions
(\ref{summary1}) above to calculate the 16 matrix elements of the
propagator.

\subsubsection{Outline of the Calculation}
\label{outline of the calculation}

Let us outline this rather cumbersome calculation in seven steps.

1. We recall the form of e.g. the massless fermionic propagator
(\ref{fermionpropmassless2}) and realise we have to extract an
operator acting on $x$, rather than on $\tilde{x}$.

The reader can easily see that the 16 matrix elements of the
propagator contain products of two Hankel functions. Some of these
products involve Hankel functions at the same order, but others
encompass Hankel functions where the order differs. Since we can
only easily integrate products of Hankel functions of equal order
we need to transform precisely the latter products by making use
of identities (\ref{Hankelprops}), e.g.:
\begin{eqnarray}\label{operatorextract}
&& H_{\nu_{+}}^{(m)}(-k\eta) H_{\nu_{-}}^{(n)}(-k\tilde{\eta}) =
\\
&& \qquad - \frac{e^{\pm
i\pi(\nu_{-}-1)}}{k}\left(\frac{\mathrm{d}}{\mathrm{d}\eta}+
\frac{\nu_{-}}{\eta}\right) H_{\nu_{-}}^{(m)}(-k\eta)
H_{\nu_{-}}^{(n)}(-k\tilde{\eta})\nonumber ,
\end{eqnarray}
where $+$ or $-$ applies when $m$ equals 1 or 2, respectively.
Clearly, the ``extracted operator'' is a function of $\eta$.

2. We rewrite the Hankel functions in terms of MacDonald
functions. By making use of \cite{Gradshteyn,Prudnikov} we find:
\begin{subequations}
\label{HankelMacDonald}
\begin{eqnarray}
H_{\nu}^{(1)}(z) &=& - \frac{2i}{\pi}e^{-i\frac{\pi}{2}\nu} K_{\nu}(-iz) \label{HankelMacDonalda}\\
H_{\nu}^{(2)}(z) &=& \frac{2i}{\pi}e^{i\frac{\pi}{2}\nu}
K_{\nu}(iz)  \label{HankelMacDonaldb}\,,
\end{eqnarray}
\end{subequations}
where $K_{\nu} (z)$ is a MacDonald function.

3. The emerging matrix structure of equation
(\ref{Feynmanpropagator2}) is most easily uncovered by evaluating
its $2\times 2$-matrix constituents successively and writing the
result in terms of Pauli matrices. Recalling equations
(\ref{kvector1}) and (\ref{kvector2}) we derive:
\begin{equation}\label{convertingderivatives}
k \left(\hat{k}_{x}\sigma_{x} + \hat{k}_{y}\sigma_{y} +
\hat{k}_{z}\sigma_{z} \right) = k_{j} \sigma^{j} \rightarrow \pm i
\sigma^{j}
\partial_{j} \,,
\end{equation}
where the $-$ or $+$ sign applies depending on whether one deals
with the $\theta (\eta-\tilde{\eta})$ or $\theta
(\tilde{\eta}-\eta)$ contribution in (\ref{Feynmanpropagator2}),
respectively. The last step is possible because only the exponent
depends on $\mathbf{x}$. Note finally that the index $j$ in
equation (\ref{convertingderivatives}) is supposed to run over
(flat) spatial indices exclusively.

4. The $\mathbf{k}$ independent contribution to the propagator can
thus be pulled out of the Fourier integral. These matrices
simplify considerably when we consider the following identities:
\begin{eqnarray}
&& \left(i\gamma^{b}\partial_{b}+am\right)\frac{1 \pm
\gamma^{0}}{2}
\label{matrixstructure} \\
&& \qquad =  \frac{1}{2}
\begin{pmatrix}
\pm i\partial_{\eta} \pm i\sigma^{j}\partial_{j}+am &
i\partial_{\eta}+i\sigma^{j}\partial_{j} \pm am \\
i\partial_{\eta}-i\sigma^{j}\partial_{j} \pm am &
\pm i\partial_{\eta} \mp i\sigma^{j}\partial_{j}+am \\
\end{pmatrix} \,.
\nonumber
\end{eqnarray}
Clearly, we recognise the complex conjugate of the Dirac operator
and the $(1\pm\gamma^{0})/2$ structure we are familiar with from
the de Sitter propagator.

5. Let us present an intermediate result that already captures
much of the final structure of our propagator. For brevity, let us
only consider the contribution to the propagator arising from the
first $\theta$-function in the propagator, i.e.: when
$\eta>\tilde{\eta}$. This part of the Feynman propagator reads:
\begin{eqnarray}
iS_{\mathrm{F}}^{ab} (x,\tilde{x})\Big |_{\eta > \tilde{\eta}} &=&
a^{-\frac{D-1}{2}} \tilde{a}^{-\frac{D-1}{2}}
(i\gamma^{b}\partial_{b} + am) \label{intermediateresult1}
\\
&& \times \Big[ \frac{\sqrt{\eta\tilde{\eta}}}{\pi} \Big\{
\frac{1+\gamma^{0}}{2} \! \int \!
\frac{\mathrm{d}^{\scriptscriptstyle{D-1}}\mathbf{k}}{(2\pi)^{\scriptscriptstyle{D-1}}}
K_{-}(k,\eta,\tilde{\eta})
\nonumber \\
&& \quad + \frac{1-\gamma^{0}}{2} \int
\frac{\mathrm{d}^{\scriptscriptstyle{D-1}}\mathbf{k}}{(2\pi)^{\scriptscriptstyle{D-1}}}
K_{+}(k,\eta,\tilde{\eta}) \Big\} \Big] \nonumber ,
\end{eqnarray}
where:
\begin{subequations}
\begin{eqnarray}
K_{-}(k,\eta,\tilde{\eta}) &=& |c|^{2} \, K_{\nu_{-}}(ik\eta)
K_{\nu_{-}}(-ik\tilde{\eta}) \label{intermediateresult2} \\
&& - ic d^{\ast} \, K_{\nu_{-}}(ik\eta)
K_{\nu_{-}}(ik\tilde{\eta}) \nonumber \\
&& - ic^{\ast} d \, K_{\nu_{-}}(-ik\eta)
K_{\nu_{-}}(-ik\tilde{\eta}) \nonumber \\
&& - |d|^{2} \, K_{\nu_{-}}(-ik\eta) K_{\nu_{-}}(ik\tilde{\eta})
\nonumber \,,
\end{eqnarray}
and where:
\begin{eqnarray}
K_{+}(k,\eta,\tilde{\eta}) &=& |c|^{2} \, K_{\nu_{+}}(ik\eta)
K_{\nu_{+}}(-ik\tilde{\eta}) \\
&& + ic d^{\ast} \, K_{\nu_{+}}(ik\eta)
K_{\nu_{+}}(ik\tilde{\eta}) \nonumber \\
&& + ic^{\ast} d \, K_{\nu_{+}}(-ik\eta)
K_{\nu_{+}}(-ik\tilde{\eta}) \nonumber \\
&& - |d|^{2}\,  K_{\nu_{+}}(-ik\eta) K_{\nu_{+}}(ik\tilde{\eta})
\nonumber \,.
\end{eqnarray}
\end{subequations}
The contribution to the propagator when $\eta < \tilde{\eta}$ is
given by the same expression, where we only need to replace the
normalisation constants $c$ and $d$ by $f$ and $g$, respectively.

The complicated structure of the propagator in spinor space has
dramatically simplified by extracting the appropriate operators in
position space, which could already be expected from the
propagator in de Sitter spacetime.

6. We need to perform the Fourier integrals over the functions
$K_{-}(k,\eta,\tilde{\eta}) $ and $ K_{+}(k,\eta,\tilde{\eta}) $
next. We use:
\begin{eqnarray}\label{spericalsymmetry}
&&
\int\frac{\mathrm{d}^{\scriptscriptstyle{D-1}}\mathbf{k}}{(2\pi)^{\scriptscriptstyle{D-1}}}
e^{i\mathbf{k}\cdot
\mathbf{x}} f(k) \\
&& \qquad = \frac{2}{(4\pi)^{\frac{\scriptscriptstyle{D-1}}{2}}}
\int_{0}^{\infty} \mathrm{d}k \, k^{\scriptscriptstyle{D-2}}
\frac{J_{\frac{\scriptscriptstyle{D-3}}{2}}(k x)}{(\frac{1}{2}k x
)^{\frac{\scriptscriptstyle{D-3}}{2}}} f(k) \nonumber\,,
\end{eqnarray}
which is valid for any function $f(k)$ that depends solely on
$k=\|\mathbf{k}\|$. Here, $ J_{\mu}(k x)$ is a Bessel function of
the first kind. The reader can easily verify that all the Fourier
integrals we need to perform are of the following form:
\begin{equation}\label{Fourierintegral}
\int_{0}^{\infty} \mathrm{d}k \, k^{\mu+1} J_{\mu}(c k) K_{\nu} (
\alpha k) K_{\nu} (\beta k)\,,
\end{equation}
where the arguments of the MacDonald functions are purely
imaginary:
\begin{subequations}
\label{Fourierintegral2}
\begin{eqnarray}
\alpha & =& \pm i \eta \label{Fourierintegral2a} \\
\beta &=& \pm i \tilde{\eta} \label{Fourierintegral2b} \,.
\end{eqnarray}
\end{subequations}
We perform these integrals in appendix \ref{Fourier Transforming
Hankel Functions} by making use of \cite{Gradshteyn}. The result
of this integral can be expressed in terms of the Gauss'
hypergeometric function $\phantom{.}_{2}F_{1}$. The
$\theta$-functions in the propagator in combination with the
arguments of the hypergeometric functions allow us to rewrite
these arguments in terms of the $y$-functions (\ref{y}) we
introduced in section \ref{Properties of FLRW spacetimes}.

A well-known fact in Minkowski or de Sitter spacetime extends to
our analysis in FLRW spacetimes: the small real contribution
needed to make the Fourier integrals convergent determines the
$\varepsilon$ pole prescription of the various contributions to
the propagator.

\begin{widetext}
7. The final result for the constant $\epsilon$ and constant
$\zeta$ FLRW fermion propagator reads:
\begin{equation}\label{fermionpropagator}
i S_{F}^{ab}(x,\tilde{x}) = a \left( i \gamma^{\mu}\nabla_{\mu} +
m\right) \frac{(a\eta\,\tilde{a}
\tilde{\eta})^{-\frac{D-2}{2}}}{\sqrt{a \tilde{a}}} \left[i S_{+}
(x,\tilde{x}) \frac{1+ \gamma^{0}}{2} + i S_{-} (x,\tilde{x})
\frac{1- \gamma^{0}}{2} \right] \,,
\end{equation}
where:
\begin{eqnarray}\label{fermionpropagator2}
i S_{\pm}(x,\tilde{x}) &=& \frac{\Gamma\left(\frac{D}{2} \pm i
\zeta \right) \Gamma \left(\frac{D-2}{2} \mp i \zeta\right)}
{\left(4\pi\right)^{\scriptscriptstyle{D/2}}
\Gamma\left(\frac{D}{2}\right)}\Bigg\{ |c|^{2}
\phantom{1}_{2}F_{1} \left(\frac{D}{2} \pm i \zeta ,\frac{D-2}{2}
\mp i \zeta ;\frac{D}{2}
; 1-\frac{y_{++}(x;\tilde{x})}{4} \right) \\
&& \qquad\qquad\qquad\qquad\qquad\qquad \mp ic d^{\ast} \,\,\,\,
e^{i\pi \frac{D-1}{2}} \phantom{1}_{2}F_{1} \left(\frac{D}{2} \pm
i \zeta ,\frac{D-2}{2} \mp i \zeta ;\frac{D}{2} ;
1-\frac{y_{+-}(x;\tilde{\bar{x}})}{4} \right)
\nonumber \\
&& \qquad\qquad\qquad\qquad\qquad\qquad \mp ic^{\ast} d \,
e^{-i\pi \frac{D-1}{2}} \phantom{1}_{2}F_{1} \left(\frac{D}{2} \pm
i \zeta ,\frac{D-2}{2} \mp i \zeta
;\frac{D}{2} ; 1-\frac{y_{-+}(x;\tilde{\bar{x}})}{4} \right) \nonumber  \\
&&  \qquad\qquad\qquad\qquad\qquad\qquad -|d|^{2}
\phantom{1}_{2}F_{1} \left(\frac{D}{2} \pm i \zeta ,\frac{D-2}{2}
\mp i \zeta ;\frac{D}{2} ; 1-\frac{y_{--}(x;\tilde{x})}{4} \right)
\Bigg\} \nonumber \,,
\end{eqnarray}
where the $y$-functions are given in equation (\ref{y}) and
$\zeta$ in (\ref{zeta}). Moreover, to write the propagator in
terms of the $y$-functions we introduced the following notation:
\begin{equation}\label{antipoldalcoordinate}
\tilde{\bar{x}} = \overline{(\tilde{\eta},\tilde{\mathbf{x}})} =
(- \tilde{\eta},\tilde{\mathbf{x}})\,.
\end{equation}
\end{widetext}
One can think of $\tilde{\bar{x}}$ as an antipodal coordinate (see
e.g.: \cite{Mottola:1984ar, Allen:1985ux, Goldstein:2003ut,
Einhorn:2003xb, Janssen:2008dw}). Note that we extract an operator
in (\ref{fermionpropagator}) that does not depend on $\tilde{x}$.
However, there is nothing special about the $x$ leg of the
propagator. Keeping remark 1 above in mind, we know that we could
have equally well written the propagator (\ref{fermionpropagator})
in terms of an operator acting on $\tilde{x}$ by replacing:
\begin{equation}\label{operatorreplacement}
a \left( i \gamma^{\mu}\nabla_{\mu} + m\right) \rightarrow
\tilde{a} \left( i \tilde{\gamma}^{\mu}\tilde{\nabla}_{\mu} +
m\right) \,.
\end{equation}
This propagator would then have satisfied the Dirac equation
(\ref{fermionprop}) in terms of $\tilde{x}$.

As a first check of this result, note that if we send $m
\rightarrow 0$ and $d \rightarrow 0$, the massive fermionic FLRW
propagator (\ref{fermionpropagator}) correctly yields the massless
fermionic FLRW propagator earlier derived in equation
(\ref{fermionpropmassless2}). Again, in order to derive this
result we needed the explicit form of $\xi_{h}$ given by
(\ref{helicity1}). The operator in (\ref{fermionpropagator}) has
thus been constructed in $D=4$. However, we can analytically
continue again to arbitrary dimensions.

Finally, let us study the IR behaviour of our propagator. Note
that the coincidence propagator (in Fourier space) is IR finite
due to the Pauli exclusion principle. In position space, note that
the IR limit of all the y-functions in equation (\ref{y}) is:
\begin{equation}\label{IRlimity}
y(x;\tilde{x}) \rightarrow \pm \infty \,,
\end{equation}
where the plus or minus sign applies for infinite spacelike or
timelike separation, respectively. If we recall the following
well-known relation:
\begin{widetext}
\begin{equation}\label{Hypergeometric2F1relationIR}
\! \phantom{1}_{2}F_{1}\left (a , b; c ; z \right) =
\frac{\Gamma(c) \Gamma(b \!-\! a)}{\Gamma(b) \Gamma(c\!-\!a)}
(1\!-\!z)^{-a}  \! \phantom{1}_{2}F_{1}\! \left (\!a , c\!-\!b;
a\!-\!b+\!1 ; \frac{1}{1\!-\!z}\! \right) +\! \frac{\Gamma(c)
\Gamma(a \!-\!b)}{\Gamma(a) \Gamma(c\!- \!b)}
(1\!-\!z)^{-b}\!\phantom{1}_{2} F_{1}\!\left(\!b ,c\!-\!a;
b\!-\!a\!+\!1; \frac{1}{1\!-\!z}\! \right)\! ,
\end{equation}
we can easily extract the leading order IR behaviour of the
propagator in equation (\ref{fermionpropagator2}):
\begin{eqnarray}\label{fermionpropagator2IR}
i S_{\pm}(x,\tilde{x}) &=& \frac{\Gamma\left(\frac{D-2}{2} \mp i
\zeta \right) \Gamma \left(1 \pm 2 i \zeta\right)}
{\left(4\pi\right)^{\scriptscriptstyle{D/2}} \Gamma\left(1 \pm
i\zeta \right)}\Bigg\{ |c|^{2} \left(\frac{y_{++}(x;\tilde{x})}{4}
\right)^{-\left(\frac{D-2}{2}\pm i\zeta\right)} \mp ic d^{\ast}
\,\,\,\, e^{i\pi \frac{D-1}{2}}
\left(\frac{y_{+-}(x;\tilde{\bar{x}})}{4}
\right)^{-\left(\frac{D-2}{2}\pm i\zeta\right)} \\
&& \qquad\qquad\qquad\quad \mp ic^{\ast} d \, e^{-i\pi
\frac{D-1}{2}} \left(\frac{y_{-+}(x;\tilde{\bar{x}})}{4}
\right)^{-\left(\frac{D-2}{2}\pm i\zeta\right)} -|d|^{2}
\left(\frac{y_{--}(x;\tilde{x})}{4}
\right)^{-\left(\frac{D-2}{2}\pm i\zeta\right)} \Bigg\}
+\mathcal{O}\left(y^{\frac{D}{2}\pm i\zeta}\right) \nonumber \,,
\end{eqnarray}
\end{widetext}
Note that for the spacelike IR, all y-functions behave identically
and the result above can be simplified further. For the timelike
IR, the $\varepsilon$-prescription in equation (\ref{y}) dictates
how to take the powers of the y-functions. Of course, one has to
insert equation (\ref{fermionpropagator2IR}) above into
(\ref{fermionpropagator}) to obtain the full propagator in the IR
limit.

The fermionic propagator (\ref{fermionpropagator}) is analogous to
the scalar and graviton propagator in universes with constant
deceleration parameter $q=\epsilon-1$ derived in
\cite{Janssen:2008dp, Janssen:2008dw,Janssen:2008px}. Unlike in
the case of scalars and gravitons, where quite generically the
Bunch-Davies vacuum is IR divergent, the fermionic propagator is
IR finite due to the Pauli exclusion principle which forbids an
accumulation of fermions in the deep IR. The IR behaviour for the
scalar and graviton are important, as it can comprise secular
effects (that grow as a power in time) in for example
$V_{\mathrm{eff}}(\Phi)$ (see: \cite{Janssen:2009pb}) and
$T^{\mu\nu}_{1\,\mathrm{loop}}$ (see: \cite{Janssen:2008px,
inprep}). Mathematically speaking, Pauli blocking is enforced by
the imaginary $i$ in equation (\ref{DiracEq4}), which ultimately
carries through in the complex indices of the hypergeometric
functions in the propagator (\ref{fermionpropagator2}).

\subsection{Fermionic de Sitter Propagator}
\label{Fermionic de Sitter propagator}

Having derived the fermionic FLRW propagator, let us verify
whether it correctly reduces to the fermionic de Sitter
propagator, which is a well-known result in the literature. The
fermion propagator in de Sitter spacetime reads
\cite{Candelas:1975du, Miao:2006pn}:
\begin{eqnarray}\label{fermionpropmassdeSitter}
&& \! i S_{F}^{\Lambda}(x,\tilde{x}) = a \left( i \gamma^{\mu}\nabla_{\mu} + m\right) \\
&& \qquad \times \frac{H^{\scriptscriptstyle{D-2}}}{\sqrt{a
\tilde{a}}} \left[i S_{+}^{\Lambda} (x,\tilde{x}) \frac{1+
\gamma^{0}}{2} + i S_{-}^{\Lambda} (x,\tilde{x}) \frac{1-
\gamma^{0}}{2} \right] \nonumber,
\end{eqnarray}
where the superscript $\Lambda$ denotes de Sitter spacetime and
where:
\begin{eqnarray}\label{fermionpropmassdeSitter2}
i S_{\pm}^{\Lambda} (x,\tilde{x}) &=&
\frac{1}{\left(4\pi\right)^{\scriptscriptstyle{D/2}}}
\frac{\Gamma\left(\frac{D}{2}-1 \mp i\frac{m}{H}\right)
\Gamma\left(\frac{D}{2}\pm i\frac{m}{H}\right)}{\Gamma\left(\frac{D}{2}\right)} \\
&& \times_{2}F_{1}\left(\frac{D}{2}-1 \mp i\frac{m}{H},
\frac{D}{2} \pm i\frac{m}{H}; \frac{D}{2} ; 1-\frac{y}{4}
\right)\! . \nonumber
\end{eqnarray}
Here, $y$ is shorthand for $y_{++}(x,\tilde{x})$. If we send
$\epsilon \rightarrow 0$ and set $c=1$ and $d=0$, equation
(\ref{fermionpropagator}) above correctly reproduces the de Sitter
result. Moreover, using:
\begin{equation}\label{Hypergeometricexpansion1}
\phantom{1}_{2}F_{1}\left(\frac{D}{2}-1,\frac{D}{2}; \frac{D}{2}
;1-\frac{y}{4} \right) =
\left(\frac{y}{4}\right)^{\scriptscriptstyle{1-D/2}}\,,
\end{equation}
note that the massless limit of (\ref{fermionpropmassdeSitter})
indeed corresponds to (\ref{fermionpropmassless2}) in de Sitter
space. We conclude that the fermionic propagator in FLRW
spacetimes correctly reduces to the known cases in existing
literature.

\section{One Loop Effective Action}
\label{One Loop Effective Action}

As a simple application for the propagator derived in section
\ref{The Feynman Propagator}, we calculate the one loop
contribution to the effective action. This is important to study
the impact of fermions on the evolution of the background
spacetime and of scalar fields coupled to these fermions. We
renormalise using the minimal subtraction dimensional
renormalisation technique.

\subsection{Evaluating the One Loop Backreaction}
\label{Evaluating the One Loop Backreaction}

The one loop effective action formally reads:
\begin{eqnarray}\label{effectiveaction1}
\Gamma_{1} &=& - i \mathrm{Tr} \log \left[ \sqrt{-g}\left(i
\gamma^{\mu}
\nabla_{\mu} - m \right)\right] \\
&=&  \int^{m}\! \mathrm{d}\bar{m} \, \mathrm{Tr} \left[ \sqrt{-g}
\, i S_{\mathrm{F}}^{ab}(x,\tilde{x})\right] \nonumber\,.
\end{eqnarray}
Here, the trace is both over spatial and spinorial indices. We
have been able to evaluate the logarithm at the expense of losing
all mass independent contributions to the effective action.

The effective action for a massless fermion is completely fixed by
the trace anomaly which this field is known to exhibit (see for
example \cite{Birrell:1982ix, Antoniadis:2006wq, Koksma:2008jn}).
The trace anomaly can be generated from a finite non-local
effective action (see e.g.: \cite{Antoniadis:2006wq}). It can
alternatively be generated from an infinite but local effective
action \cite{Koksma:2008jn}. We will henceforth neglect all
contributions from the trace anomaly to the effective action in
this paper because it has already been extensively discussed in
the literature.

We treat the covariant derivative according to
(\ref{covariantderivative2}) and we can easily expand the
hypergeometric functions in equation (\ref{fermionpropagator2}) at
coincidence. In dimensional regularisation, the $D$ dependent
powers in this expansion do not contribute. These are generated
for the $y_{++}$ and $y_{--}$ contributions to
(\ref{fermionpropagator2}) by the first term in:
\begin{eqnarray}\label{Hypergeometric2F1relation}
&& \phantom{1}_{2}F_{1}\left (a , b; c ; z \right) =
\frac{\Gamma(c) \Gamma(a+b - c)}{\Gamma(a) \Gamma(b)}
(1-z)^{c-a-b} \\
&& \qquad\qquad  \times_{2}\!F_{1} \left (c-a , c-b; c -a-b+1; 1-z \right) \nonumber \\
&& \quad + \frac{\Gamma(c) \Gamma(c -a -b)}{\Gamma(c-a)
\Gamma(c-b)}\! \phantom{1}_{2}F_{1}\left (a ,b; a+b-c+1; 1-z
\right) \nonumber .
\end{eqnarray}
Moreover, when we trace over spacetime indices note there is no
contribution at all from the derivative term hitting one of the
hypergeometric functions:
\begin{equation}\label{derivativeony}
\lim_{\tilde{x}\rightarrow x} \sum_{n=0}^{\infty}
\gamma^{b}\partial_{b} \, y_{++}^{n} (x,\tilde{x}) = -
\lim_{\tilde{x}\rightarrow x} \sum_{n=0}^{\infty}  n \gamma^{0}
\frac{y_{++}^{n}(x,\tilde{x})}{\eta} = 0\,.
\end{equation}
Here, $n$ is the integer valued coefficient of the Taylor
expansion of a hypergeometric function. A similar identity holds
for $y_{--}$. Likewise, for $y_{+-}$ or $y_{-+}$, we have e.g.:
\begin{equation}\label{derivativeony2}
\lim_{\tilde{x}\rightarrow x} \gamma^{b}\partial_{b} \, y_{+-}
(x,\tilde{\bar{x}}) = 0\,.
\end{equation}
Note that both $y_{+-}$ and $y_{-+}$ depend on the antipodal
coordinate $\tilde{\bar{x}}$ in equation
(\ref{fermionpropagator}). At coincidence, we also use:
\begin{eqnarray}\label{intermediatestep}
&& (a
\tilde{a})^{-\frac{D-1}{2}}(\eta\tilde{\eta})^{-\frac{D-2}{2}}\left(
am \mp \frac{D-2}{2}\frac{i}{\eta}\right) \\ && \qquad\quad
\rightarrow \pm i\left(H|1-\epsilon|
\right)^{\scriptscriptstyle{D}-2} H(1-\epsilon)\left(
\frac{D-2}{2}\mp i\zeta\right) \nonumber \,,
\end{eqnarray}
The result is:
\begin{widetext}
\begin{eqnarray} \label{effectiveactionfullresult}
\Gamma_{1} &=& \int \mathrm{d}^{\scriptscriptstyle{D}}\! x
\sqrt{-g} \frac{(H|1-\epsilon|)^{\scriptscriptstyle{D}-2}}
{(2\pi)^{\scriptscriptstyle{D}/2}}\Bigg\{
\Gamma\left(1-\frac{\scriptstyle{D}}{2}\right)  (|c|^{2} -
|d|^{2})  \int^{m}\mathrm{d}\tilde{m} \, \tilde{m} \frac{
\Gamma\left(\frac{D}{2}+i \tilde{\zeta}\right)
\Gamma\left(\frac{D}{2}-i\tilde{\zeta}\right)
}{\Gamma\left(1-i\tilde{\zeta}\right)\Gamma\left(1+i\tilde{\zeta}\right)}
 \\
&& \qquad\qquad\qquad\qquad\quad + H(1-\epsilon) \left( c d^{\ast}
\, e^{i\pi \frac{D-1}{2}} + c^{\ast} d \, e^{-i\pi \frac{D-1}{2}}
\right) \int^{m}\mathrm{d}\tilde{m} \,
\frac{\Gamma\left(\frac{D}{2}+i \tilde{\zeta}\right)
\Gamma\left(\frac{D}{2}-i\tilde{\zeta}\right)}{\Gamma\left(\frac{D}{2}\right)}
\Bigg\} \nonumber\,.
\end{eqnarray}
\end{widetext}
Let us compare this result with the known calculations in the
literature in de Sitter spacetime ($\epsilon\rightarrow 0$) and in
the vacuum ($|d|=0$). Note the two omissions in the Candelas and
Raine effective action \cite{Candelas:1975du}. When tracing over
spinor indices, we have:
\begin{equation}\label{traceprojectors}
\mathrm{Tr}\frac{1\pm\gamma^{0}}{2} = 2^{\frac{D}{2}-1}\,,
\end{equation}
and this trace does not, as is apparent from
\cite{Candelas:1975du}, equal $D/2$ because a spinor in $D$
spacetime dimensions has $2^{\frac{D}{2}}$ degrees of freedom (see
e.g.: \cite{DeWit:1986it}). Indeed, this has already been noted by
\cite{Inagaki:1995jp, Inagaki:1997kz}. Moreover, the effective
action in \cite{Candelas:1975du} misses a factor of 2 stemming
from the separate particle and anti-particle contributions, which
has been corrected for in e.g. \cite{Miao:2006pn}.

The reader can easily verify that
(\ref{effectiveactionfullresult}), taking these errors into
account agrees with \cite{Candelas:1975du, Miao:2006pn} in de
Sitter spacetime and in the vacuum.

\subsection{Dimensional Regularisation} \label{Dimensional
Regularisation}

In the spirit of dimensional regularisation, we now expand around
$D=4$. We can use the familiar result:
\begin{equation}\label{Regularisation1}
\Gamma(x+y\tilde{\epsilon}) = \Gamma(x)\left[1+y\tilde{\epsilon}\,
\psi(x)\right] + \mathcal{O}(\tilde{\epsilon}^{2})\,,
\end{equation}
where $\tilde{\epsilon} \ll 1$ and where $\psi(x)$ is the digamma
function defined by $\psi(x) = d \log \Gamma(x) /dx$. Anticipating
the form of the counterterms we will add shortly, we introduce a
scale $\mu$ by:
\begin{eqnarray}\label{Regularisation2}
&& (H|1-\epsilon|)^{\scriptscriptstyle{D}} = (H(1-\epsilon))^{4}
\mu^{\scriptscriptstyle{D}-4} \left( \frac{H|1-\epsilon|}{\mu}
\right)^{\scriptscriptstyle{D}-4}
\\
&& \quad \simeq (H(1-\epsilon))^{4} \mu^{\scriptscriptstyle{D}-4}
\left[ 1+ \frac{D-4}{2} \ln
\left(\frac{H^{2}(1-\epsilon)^{2}}{\mu^{2}}\right)\!\right]
\nonumber ,
\end{eqnarray}
we we neglected $\mathcal{O}\{(D-4)^{2}\}$ contributions. Finally,
we recall the familiar result:
\begin{equation}\label{Regularisation3}
\Gamma\left(1-\frac{D}{2}\right) = \frac{2}{D-4} +
\gamma_{\mathrm{\scriptscriptstyle{E}}}-1 + \mathcal{O}(D-4)\,,
\end{equation}
where $\gamma_{\mathrm{\scriptscriptstyle{E}}}$ is the
Euler--Mascheroni constant. We thus arrive at:
\begin{eqnarray}
\! \Gamma_{1} &=& \int \mathrm{d}^{\scriptscriptstyle{D}}\!x
\frac{\sqrt{-g}}{4\pi^{2}} \Bigg\{ \left(|c|^{2}-
|d|^{2}\right)\mu^{\scriptscriptstyle{D}-4} \Bigg[ \nonumber \\
&&  \qquad \times\left(  \frac{1}{2} m^{2} H^{2} (1-\epsilon)^{2}
+\frac{1}{4} m^{4} \right)  \nonumber
\\
&& \qquad \times \left( \frac{2}{D-4} +
\gamma_{\mathrm{\scriptscriptstyle{E}}} -1 +
\ln\left[\frac{H^{2}(1-\epsilon)^{2}}{2\pi\mu^{2}}\right] \right)
\label{effectiveactionexpansionD4} \\
&& \! +\! \int^{m}\!\!\mathrm{d}\tilde{m}\!
\left(\tilde{m}^{3}\!+\!\tilde{m}H^{2}(1\!-\!\epsilon)^{2}
\right)\!\left[
\psi(2\!-\!i\tilde{\zeta})\!+\!\psi(2\!+\!i\tilde{\zeta})\!
\right]\!
\Bigg]\! \nonumber \\
&& \! + 2 H^{3}(1-\epsilon)^{3} \mathrm{Im}\left(c d^{\ast}\right)
\! \int^{m}\! \mathrm{d}\tilde{m} \, \Gamma(2+i \tilde{\zeta})
\Gamma(2-i\tilde{\zeta})\Bigg\} \nonumber \!.
\end{eqnarray}
Here, $\mathrm{Im}(c d^{\ast})=|c||d|\sin(\phi_{c}-\phi_{d})$
denotes the imaginary part of $c d^{\ast}$.

The one loop backreaction contains a divergence when $D=4$.
However, appreciate that the contribution to the effective action
multiplying the mixed coefficients is finite in $D=4$.

\subsection{Tree level Friedmann Equations} \label{Tree level
Friedmann Equations}

Anticipating the renormalisation procedure in the next section, we
need the tree level, i.e.: classical, Friedmann equations of
motion responsible for driving the expansion of the universe. The
system of interest is thus given by:
\begin{eqnarray}\label{actiontotalgravity}
S &=& \frac{1}{\kappa} \int \mathrm{d}^{\scriptscriptstyle{D}}\! x
\sqrt{-g}
\left\{R-(D-2)\Lambda\right\} \\
&& + \int \mathrm{d}^{\scriptscriptstyle{D}}\! x \sqrt{-g} \left\{
-\frac{1}{2}g^{\mu\nu}\partial_{\mu} \varphi \partial_{\nu}
\varphi -V(\varphi) \right\} \nonumber \\
&& + \int \mathrm{d}^{\scriptscriptstyle{D}}\! x \sqrt{-g}\left\{
-\frac{1}{2}g^{\mu\nu}\partial_{\mu} \phi \partial_{\nu} \phi
-V(\phi)- g_{\mathrm{\scriptscriptstyle{Y}}} \bar{\psi}\psi\phi \right\} \nonumber \\
&& + \int \mathrm{d}^{\scriptscriptstyle{D}}\! x
\sqrt{-g}\left\{\frac{i}{2}\left[\bar{\psi} \gamma^{\mu}
\nabla_{\mu} \psi - \left (\nabla_{\mu} \bar{\psi}\right)
\gamma^{\mu} \psi\right] \right\} \,. \nonumber
\end{eqnarray}
Here, $\kappa = 16\pi \mathrm{G_{N}}$ represents the rescaled
Newton constant and $\Lambda$ denotes the cosmological constant.
The reader can easily recognise the usual Einstein-Hilbert action
on the first line. We introduce a new scalar field $\varphi$,
responsible for the dynamics of the universe. By an appropriate
choice of the potential $V(\varphi)$, a scalar field can mimic any
mixture of fluids relevant for the evolution of our universe (see
e.g. \cite{Padmanabhan:2002cp}). The last two lines in equation
(\ref{actiontotalgravity}) above are identical to
(\ref{actiontotal}) and contain the fermion field and the scalar
field $\phi$ generating the mass of the fermion through a Yukawa
coupling, as in (\ref{fermionmass}). We write:
\begin{subequations}
\label{backgroundfields1}
\begin{eqnarray}
\varphi(x)&=& \varphi_{0}(t)+\delta\varphi(x) \label{backgroundfields1a}\\
\phi(x)&=& \phi_{0}(t)+\delta\phi(x)\,, \label{backgroundfields1b}
\end{eqnarray}
\end{subequations}
where we assume the background fields $\varphi_{0}(t)$ and
$\phi_{0}(t)$ to be homogeneous. Moreover, we assume that
\mbox{$\rho_{\varphi} \gg \rho_{\phi}$}, such that $\varphi(x)$
drives the dynamics of the universe as stated above. The classical
Friedmann equations of motion are:
\begin{subequations}
\label{treelevelFriedmann1}
\begin{eqnarray}
H^{2}-\frac{1}{D-1}\Lambda - \frac{\kappa
\left(\frac{1}{2}\dot{\varphi}_{0}^{2}+V(\varphi_{0})\right)
}{(D-1)(D-2)}
&=& 0 \phantom{11} \label{treelevelFriedmann1a}\\
\dot{H}+\frac{D-1}{2}H^{2}-\frac{\Lambda}{2} + \frac{\kappa
\left(\frac{1}{2}\dot{\varphi}_{0}^{2}-V(\varphi_{0})\right)
}{2(D-2)}
 &=& 0\,,\phantom{11}
\label{treelevelFriedmann1b}
\end{eqnarray}
and moreover, the classical scalar field equation of motion reads:
\begin{equation}\label{treelevelFriedmann1c}
\ddot{\varphi}_{0}+(D-1)H\dot{\varphi}_{0}+\frac{\partial
V}{\partial \varphi} (\varphi_{0}) = 0 \,.
\end{equation}
\end{subequations}
We can thus derive the following identities:
\begin{subequations}
\label{treelevelFriedmann2}
\begin{eqnarray}
\sqrt{\kappa}\, \dot{\varphi}_{0} &=& \sqrt{2(D-2)\epsilon} \, H
\label{treelevelFriedmann2a}\\
\sqrt{\kappa} \, \frac{\partial V}{\partial \varphi}(\varphi_{0})
&=& - \sqrt{2(D-2) \epsilon}\, (D-1-\epsilon)H^{2}
\label{treelevelFriedmann2b}\\
\sqrt{\kappa} \, \ddot{\varphi}_{0} &=& - \sqrt{2(D-2) \epsilon}
\,
\epsilon H^{2} \label{treelevelFriedmann2c}\\
\frac{\partial^{2} V}{\partial \varphi^{2}}(\varphi_{0}) &=&
2(D-1-\epsilon) \epsilon H^{2} \label{treelevelFriedmann2d}\,.
\end{eqnarray}
\end{subequations}
Here, we have used assumption (\ref{epsilon}).

\subsection{Renormalisation}
\label{Renormalisation}

In order to derive equation (\ref{effectiveactionexpansionD4}), we
have assumed a constant deceleration $\epsilon$, as in equation
(\ref{epsilon}). We now promote this constant to a dynamical
quantity:
\begin{equation}\label{epsilonpromotion}
\epsilon \rightarrow \epsilon(t) \,.
\end{equation}
We will motivate this step shortly. The effective action
(\ref{effectiveactionexpansionD4}) now contains divergences in
both $H(t)$ and $\epsilon(t)$ that we ought to cancel by an
appropriate counterterm action. It is not possible to identify
local covariant counterterms of curvature invariants only that
remove the singularities in both of these quantities at the level
of the effective action. We can use the tree level equations of
motion derived above in (\ref{treelevelFriedmann2}) to renormalise
the effective action, which is first order in $\hbar$. By making
use of an infinitesimal field redefinition, one can show the
following: substituting the zeroth order equation of motion into
the first order contribution to the action yields, up to one loop,
the same equation of motion as directly varying the action. We
thus add the following counterterm action:
\begin{eqnarray}\label{countertermsaction}
\Gamma_{\mathrm{ct}} &=& \int \mathrm{d}^{\scriptscriptstyle{D}}\!
x \sqrt{-g} \Big(c_{1} \phi^{2} R +
c_{2}\phi^{2}\frac{\partial^{2}V}{\partial\varphi^{2}}(\varphi_{0})
\\
&&  \phantom{\sqrt{-g}1} - c_{3}\kappa \phi^{2}
g^{\mu\nu}\partial_{\mu}\varphi
\partial_{\nu}\varphi + c_{4} \phi^{4} \Big) \nonumber \,.
\end{eqnarray}
The form of the counterterm action is unique, which can be shown
by making use of dimensional analysis and by requiring that only
two time derivatives can act on $\varphi_{0}(t)$ or $a(t)$ for
stability. These conditions limit us to the counterterms above and
possibly $\phi^{2}\varphi\Box\varphi$. However, one can easily
verify that the latter term does not have the correct form
required to cancel the divergences in
(\ref{effectiveactionexpansionD4}).

We recall relation (\ref{fermionmass}). We need to expand the
various terms in the counterterm action above around $D=4$, by
making use of $R=(D-1)(D-2\epsilon)H^{2}$ and equation
(\ref{treelevelFriedmann2}):
\begin{subequations}
\label{Counterterms1}
\begin{eqnarray}
R &\simeq & 6(2-\epsilon)H^{2} +(D-4)(7-2\epsilon)H^{2} \phantom{11111} \label{Counterterms1a} \\
\frac{\partial^{2} V}{\partial \varphi^{2}}(\varphi_{0}) &\simeq &
2 (3-\epsilon) \epsilon H^{2} + (D-4)2\epsilon H^{2}  \label{Counterterms1b} \\
\kappa g^{\mu\nu}\partial_{\mu}\varphi
\partial_{\nu}\varphi &\simeq & -4 \epsilon H^{2} - (D-4) 2\epsilon H^{2}
\label{Counterterms1c} \,,
\end{eqnarray}
\end{subequations}
where we neglected all $\mathcal{O}\{(D-4)^2\}$ contributions in
the equations above. We can now easily solve for the coefficients
in the counterterm action:
\begin{subequations}
\label{Counterterms2}
\begin{eqnarray}
c_{1} &=& -  g_{\mathrm{\scriptscriptstyle{Y}}}^{2}
\frac{|c|^{2}-|d|^{2}}{48\pi^{2}}
\frac{\mu^{\scriptscriptstyle{D}-4}}{D-4} + c_{1}^{\mathrm{f}}
 \label{Counterterms2a} \\
c_{2} &=& \phantom{-} g_{\mathrm{\scriptscriptstyle{Y}}}^{2}
\frac{|c|^{2}-|d|^{2}}{8\pi^{2}}
\frac{\mu^{\scriptscriptstyle{D}-4}}{D-4} + c_{2}^{\mathrm{f}} \label{Counterterms2b} \\
c_{3} &=& -  3 g_{\mathrm{\scriptscriptstyle{Y}}}^{2}
\frac{|c|^{2}-|d|^{2}}{32 \pi^{2}}
\frac{\mu^{\scriptscriptstyle{D}-4}}{D-4} + c_{3}^{\mathrm{f}}
\label{Counterterms2c} \\
c_{4} &=& -  g_{\mathrm{\scriptscriptstyle{Y}}}^{2}
\frac{|c|^{2}-|d|^{2}}{8 \pi^{2}}
\frac{\mu^{\scriptscriptstyle{D}-4}}{D-4} + c_{4}^{\mathrm{f}}
\label{Counterterms2d} \,,
\end{eqnarray}
\end{subequations}
where the divergent coefficients in $D=4$ are fixed to cancel the
divergences occurring in (\ref{effectiveactionexpansionD4}), and
where e.g. $c_{1}^{\mathrm{f}}$ is a finite but arbitrary
coefficient of the counterterm action. The renormalised one loop
effective action thus reads:
\begin{widetext}
\begin{eqnarray}
\Gamma_{1, \mathrm{ren}} &=& \int \mathrm{d}^{4}\!x
\frac{\sqrt{-g}}{4\pi^{2}} \Bigg\{ (|c|^{2}- |d|^{2}) \Bigg[
\left( \frac{1}{2} m^{2} H^{2} (1-\epsilon)^{2} +\frac{1}{4} m^{4}
\right) \left( \gamma_{\mathrm{\scriptscriptstyle{E}}} -1 +
\ln\left[\frac{H^{2}(1-\epsilon)^{2}}{2\pi\mu^{2}}\right] \right)
- \frac{1}{12}m^{2}H^{2}(7-5\epsilon)
 \nonumber \\
&& \qquad\qquad\qquad\qquad\qquad\quad  +
\int^{m}\!\mathrm{d}\tilde{m} \left(\tilde{m}H^{2}(1-\epsilon)^{2}
+ \tilde{m}^{3} \right) \left[
\psi(2-i\tilde{\zeta})+\psi(2+i\tilde{\zeta}) \right]
\Bigg]  \label{effectiveactionrenormalised} \\
&& \qquad\qquad\quad + 2 H^{3}(1-\epsilon)^{3} \mathrm{Im} \left(
c d^{\ast} \right) \int^{m}\! \mathrm{d}\tilde{m} \, \Gamma(2+i
\tilde{\zeta})
\Gamma(2-i\tilde{\zeta})\Bigg\} \nonumber \\
&& + \int \mathrm{d}^{4}\!x \sqrt{-g} \Big(c_{1}^{\mathrm{f}}
\phi^{2} R + c_{2}^{\mathrm{f}} \phi^{2}\frac{\partial^{2}V}
{\partial\varphi^{2}}(\varphi_{0}) - c_{3}^{\mathrm{f}} \kappa
\phi^{2} g^{\mu\nu}\partial_{\mu}\varphi
\partial_{\nu}\varphi + c_{4}^{\mathrm{f}} \phi^{4} \Big)
\nonumber\,.
\end{eqnarray}
\end{widetext}
The effective action above comprises the one loop backreaction of
fermions. This term gives rise to additional physical corrections
to the (classical) Friedmann equations, thus obtaining the
so-called quantum corrected Friedmann equations. We will study its
effect on the evolution of our universe and its impact on the
evolution of the scalar field to which the fermions are coupled in
a future publication. In this equation we have neglected all
contributions to the one loop effective action arising from the
trace anomaly.

If we expand equation (\ref{effectiveactionrenormalised}) to first
order in $\epsilon$, we can find out how large the first order
correction to the de Sitter result is. We have:
\begin{widetext}
\begin{eqnarray}
\Gamma_{1, \mathrm{ren}} &=& \Gamma_{1, \mathrm{ren}}^{\Lambda}+
\epsilon \int \mathrm{d}^{4}\!x \frac{\sqrt{-g}}{4\pi^{2}} \Bigg\{
(|c|^{2}- |d|^{2}) \Bigg[
m^{2}H^{2}\left\{\frac{5}{12}-\gamma_{\mathrm{E}}-\ln\left[\frac{H^{2}}{2\pi\mu^{2}}
\right]\right\} - \frac{1}{2}m^{4} \label{effectiveactionrenormalisedfirstorder} \\
&& \quad + \int^{m}\!\mathrm{d}\tilde{m} i \left(\tilde{m}^{2}H +
\frac{\tilde{m}^{4}}{H} \right) \Big[
\psi'(2+i\tilde{m}/H)-\psi'(2-i\tilde{m}/H)
\Big]-2\tilde{m}H^{2}\Big[
\psi(2+i\tilde{m}/H)+\psi(2-i\tilde{m}/H) \Big]
\Bigg]   \nonumber \\
&& \quad + 2 H^{3} \mathrm{Im} \left( c d^{\ast} \right)
\int^{m}\! \mathrm{d}\tilde{m} \, \Gamma(2+i \tilde{m}/H)
\Gamma(2-i\tilde{m}/H) \left[i\frac{\tilde{m}}{H}\Big(
\psi(2+i\tilde{m}/H)-\psi(2-i\tilde{m}/H) \Big)-3 \right] \Bigg\}
\nonumber + \mathcal{O}\left(\epsilon^{2}\right),
\end{eqnarray}
\end{widetext}
where the prime denotes a derivative with respect to the argument,
and the superscript $\Lambda$ represent the well-known de Sitter
result, which trivially follows from
(\ref{effectiveactionrenormalised}) by letting
$\epsilon\rightarrow 0$.

Let us at this point however already consider the $\zeta \gg 1$
limit, to extract the physical behaviour of the effective
potential resulting from (\ref{effectiveactionrenormalised}). Note
firstly that:
\begin{equation}\label{effectivepotential1}
\Gamma_{1, \mathrm{ren}} = \int \mathrm{d}^{4}\!x \sqrt{-g}
\left(-V_{\mathrm{eff}}(\phi)\right) \,,
\end{equation}
where $V_{\mathrm{eff}}(\phi)$ denotes the one loop effective
potential for $\phi$. In order to evaluate the integrals over
$\tilde{m}$ occurring in equation
(\ref{effectiveactionrenormalised}), we need the following
asymptotic expansions ($\zeta \in \mathbb{R}$):
\begin{eqnarray}\label{limitbigmass1}
\lim_{\zeta\rightarrow \infty} \Gamma(2+i\zeta)\Gamma(2-i\zeta)
&=& 2\pi |\zeta|^{3} e^{- \pi|\zeta|} \\
&& \,\,\, \times \! \left( 1+
\frac{1}{\zeta^{2}}+\mathcal{O}\left(\frac{1}{\zeta^{4}}\right)\right)
\nonumber  ,
\end{eqnarray}
and also:
\begin{eqnarray}\label{limitbigmass2}
\lim_{\zeta\rightarrow \infty} \left[
\psi(2+i\zeta)+\psi(2-i\zeta)\right] &=& \ln [\zeta^{2}] +
\frac{13}{6}\frac{1}{\zeta^{2}} \\
&& - \frac{119}{60}\frac{1}{\zeta^{4}}
+\mathcal{O}\left(\frac{1}{\zeta^{6}}\right) \nonumber \,.
\end{eqnarray}
Keeping the leading order terms of the effective potential
(\ref{effectivepotential1}) in this limit only, we find from
(\ref{effectiveactionrenormalised}) and
(\ref{limitbigmass1}--\ref{limitbigmass2}):
\begin{eqnarray}\label{effectivepotential2}
V_{\mathrm{eff}}(\phi) &\rightarrow& \!- \frac{|c|^{2}-
|d|^{2}}{16 \pi^{2 }} g^{4}_{\mathrm{\scriptscriptstyle{Y}}}
\phi^{4}
\ln\left[\frac{g_{\mathrm{\scriptscriptstyle{Y}}}^{2} \phi^{2}}{2\pi\mu^{2}}\right] \\
&&  \!+  \frac{\mathrm{Im}(c d^{\ast})}{\pi^{2}}
 g_{\mathrm{\scriptscriptstyle{Y}}}^{3} \phi^{3} H(1\!-\!\epsilon) \exp\left[{-\pi \frac{g_{\mathrm{\scriptscriptstyle{Y}}}
|\phi|} {H|1-\epsilon|}}\right] \nonumber \! ,
\end{eqnarray}
where equation (\ref{fermionmass}) has been used. Without studying
the dynamics resulting from this asymptotic effective potential,
let us briefly consider its stability properties. Clearly, this
potential is unstable in the vacuum ($|d|=0$) because the
effective potential is unbounded from below in this case. This
wild behaviour has already been recognised in de Sitter spacetime
\cite{Miao:2006pn}.

An interesting phenomenon can be observed precisely when
$|c|^2=|d|^{2} = 1/2$. This is realised for example in a thermal
state with a temperature $T$ much larger than the UV cutoff of the
theory. In this case, the asymptotic effective potential is
exponentially suppressed and its (small) contributions can either
be positive or negative.

\subsection{Discussion of the Renormalised Effective Action}
\label{Discussion of the Renormalised Effective Action}

Let us make a few general remarks. Consider quantum field theories
in de Sitter spacetime. The Hubble parameter $H$ in de Sitter
spacetime is a constant. An effective action for fermions or
scalars in de Sitter spacetime contains divergences for example
proportional to a power of $H$ (see e.g.: \cite{Candelas:1975du,
Birrell:1982ix, Inagaki:1997kz, Miao:2006pn}). The same statement
holds for divergences in the stress energy tensor in for example
the trace anomaly literature (see e.g.: \cite{Birrell:1982ix,
Antoniadis:2006wq}). At this point, one can interpret the results
differently:

\begin{enumerate}

\item The Hubble parameter can quite conservatively be treated as a
\textit{constant}. The argument for this case is simple:
throughout the calculation one simply assumed a constant Hubble
parameter.

\item The Hubble parameter is promoted to a \textit{dynamical}
quantity:
\begin{equation}\label{Hubblepromotion}
H \rightarrow H(t) \,.
\end{equation}
A reason for supporting this case would be the following: when
$\epsilon \ll 1$ one would find the de Sitter result as a leading
order contribution in each time interval where the Hubble
parameter varies only adiabatically slowly, a spacetime also known
as locally de Sitter spacetime.
\end{enumerate}

Depending on which of the two interpretations one adheres to, one
would renormalise differently. If one advocates the first point of
view, all terms of e.g. an effective action merely contribute to
the renormalised cosmological constant. If, however, one prefers
the second interpretation, one adds (dynamical) curvature
invariants in order to cancel these divergences in a covariant
manner.

We observe the following. In the effective action
(\ref{effectiveactionrenormalised}) above, the Hubble parameter is
a dynamical quantity. Hence, covariant counterterms added to
cancel UV divergences can and should be treated dynamically. This
allows us to study the backreaction of fermions in a dynamical
manner.

Albeit strictly speaking not allowed to relax the assumption
$H=\mathrm{const}$ in de Sitter spacetime, we conclude that
promoting the Hubble parameter to a dynamical quantity $H=H(t)$
has been a useful approach. When in our calculation we keep
$\epsilon$ constant as assumed, result
(\ref{effectiveactionfullresult}) favours the second
interpretation above.

Let us now return to our calculation to see what the above
considerations can tell us about our effective action
(\ref{effectiveactionfullresult}). We face the following two
options:

\begin{enumerate}

\item We treat $\epsilon$ as a \textit{constant}.

\item We promote $\epsilon$ to a \textit{dynamical}
quantity, as already stated in (\ref{epsilonpromotion}). We could
call this a spacetime with an almost constant deceleration.

\end{enumerate}

Despite the fact that throughout the calculation we have assumed
that $\epsilon$ is a constant, it will come as no surprise that we
argue in favour of the latter option for two reasons: if one would
allow $\epsilon$ to change, one would probably find our main
result for the effective action before renormalisation
(\ref{effectiveactionfullresult}) in each time interval where
$\epsilon$ changes only very slowly. Secondly, the latter option
is the generalisation of the confirmed de Sitter case above.

An obvious disadvantage of our choice is that we can only
renormalise on-shell. Consider the divergences in
(\ref{effectiveactionexpansionD4}). As already mentioned, it is
not possible to identify local covariant counterterms of curvature
invariants only that remove all singularities in $H(t)$ and
$\epsilon(t)$ at the level of the effective action. Therefore, we
relied on an on-shell renormalisation technique and inserted the
zeroth order (tree level) Friedmann equations of motion into the
effective action at first order in $\hbar$.

Finally, let us try to comfort the reader who would rather keep
$\epsilon$ constant, but who would simultaneously like to
understand how to improve on his final result for the renormalised
effective action. In this case, the only appropriate counterterms
to renormalise (\ref{effectiveactionexpansionD4}) available to us
are the following:
\begin{equation}\label{alternativerenormalisation}
\Gamma_{\mathrm{ct}}^{\mathrm{alt}} = \sqrt{-g}\left( \alpha
\phi^{2} R + \beta \phi^{4} \right)\,.
\end{equation}
We can then renormalise the theory as usual\footnote{The case
$\epsilon=2$ would have to be considered separately, for $R=0$ in
a radiation dominated universe.}, with:
\begin{subequations}
\label{countertermscoefficientsalt}
\begin{eqnarray}
\alpha &=& - g_{\mathrm{\scriptscriptstyle{Y}}}^{2}
\frac{|c|^{2}-|d|^{2}}{24\pi^{2}}
\frac{\mu^{\scriptscriptstyle{D}-4}}{D-4}
\frac{(1-\epsilon)^{2}}{2-\epsilon} + \alpha_{\mathrm{f}}
\label{countertermscoefficientsalta}
\\
\beta &=& - g_{\mathrm{\scriptscriptstyle{Y}}}^{4}
\frac{|c|^{2}-|d|^{2}}{8\pi^{2}}
\frac{\mu^{\scriptscriptstyle{D-4}}}{D-4} + \beta_{\mathrm{f}}
\label{countertermscoefficientsaltb} \,.
\end{eqnarray}
\end{subequations}
We could now study mode mixing\footnote{As already formulated in
\cite{Tsamis:2002qk} a (numerical) study of mode mixing is the
answer to the fundamental problem that we can only solve for the
mode functions exactly for a limited number of choices of $a(t)$.
In their paper, Tsamis and Woodard study mode mixing for a
massless minimally coupled scalar field in constant $\epsilon$
FLRW spacetimes. The basic strategy is as follows: in each --
possibly infinitesimally small -- time interval $\epsilon$ can be
well approximated by a constant. The mode functions in two
adjacent time intervals can be related by imposing two conditions:
(i) continuity of the mode functions and (ii) continuity of their
first derivatives. Hence, the mode functions at some later time
$\eta$ can be obtained from the initial condition at $\eta'$ by
means of a transfer matrix.}. One could simply allow $\epsilon$
and $m/H$ to vary in time and choose (if necessary infinitesimally
small) time intervals in which $\epsilon$ and $m/H$ vary only
slowly. Subsequently, one can match the solutions in two
neighbouring intervals and study mode mixing in the spirit of
\cite{Tsamis:2002qk}. In this case, this procedure would imply
that we would have to renormalise for each time interval
separately, for the coefficients in equation
(\ref{countertermscoefficientsalt}) depend on $\epsilon$. This is
not surprising: a sudden jump, even infinitesimally small, in for
example $m/H$ is not a physical phenomenon and would call for
non-local counterterms (see e.g.: \cite{Vilenkin:1982wt}).

\section{Conclusion}
\label{Conclusion}

In this paper, we have constructed the fermionic propagator in
realistic FLRW spacetimes with constant deceleration parameter
$q=\epsilon-1$. Moreover, we assumed that $m/H$ is also constant,
which can be realised in Yukawa theory by means of a scalar field
for which $\phi \propto H$. We have derived the propagator both in
the (trivial) massless case (\ref{fermionpropmassless2}) and in
the massive case (\ref{fermionpropagator}).

Two pillars support our derivation. We split the fermionic degrees
of freedom into a direct product of chirality and helicity
eigenspinors. We normalise the spinors using a consistent
canonical quantisation and we require orthogonality of particle
and anti-particle spinors.

Moreover, we verify that upon sending $\epsilon$ to zero, we
recover the known vacuum-to-vacuum de Sitter results in the
literature.

We calculate the one loop effective action induced by fermions
using our propagator. The one loop backreaction arises from
integrating out a free, quadratic fermion field and, using
dimensional regularisation, this generates a correction to the
(classical) Friedmann equations. This effective action allows for
the first time for a dynamical interpretation of $H$ such that the
study of dynamical backreaction either on the background spacetime
or on scalar fields coupled to these fermions is within reach.

\

\noindent \textbf{Acknowledgements}

\noindent JFK thanks Gerben Stavenga, Shun-Pei Miao and Timothy
Budd for useful suggestions. The authors thank Gerard 't Hooft,
Gleb Arutyunov and Tomas Janssen and acknowledge financial support
from FOM grant 07PR2522 and by Utrecht University.

\appendix

\section{Properties of Hankel Functions}
\label{Properties of Hankel Functions}

In section \ref{Particle Mode Functions} we used the following
well-known identities involving Hankel functions:
\begin{subequations}
\label{Hankelprops}
\begin{eqnarray}
H^{(1)}_{-\nu}(z) &=& e^{i\pi\nu} H^{(1)}_{\nu}(z)
\label{Hankelprops1}  \\
H^{(2)}_{-\nu}(z) &=& e^{-i\pi\nu} H^{(2)}_{\nu}(z)
\label{Hankelprops2} \\
\{ H^{(1)}_{\nu}(z) \}^{\ast} &=& H^{(2)}_{\nu^{\ast}}(z^{\ast})
\label{Hankelprops3} \\
H^{(1)}_{\nu}\left(e^{i\pi}z\right) &=& - H^{(2)}_{- \nu}(z) =
-e^{-i\pi\nu} H^{(2)}_{\nu}(z) \label{Hankelprops8} \phantom{111}\\
H^{(2)}_{\nu}\left(e^{-i\pi}z\right) &=& - H^{(1)}_{- \nu}(z) =
-e^{i\pi\nu} H^{(1)}_{\nu}(z) \,.\label{Hankelprops9}
\end{eqnarray}
The Wronskian of two Hankel functions reads:
\begin{eqnarray}
W[ H^{(1)}_{\nu}(z), H^{(2)}_{\nu}(z)] &=& -\frac{4 i}{\pi z}
 \label{Hankelprops5}\,.
\end{eqnarray}
Moreover, we made use of the following recurrence relation:
\begin{eqnarray}
H^{(i)}_{\nu-1}(z) &=& \frac{\mathrm{d}}{\mathrm{d}z}\,
H^{(i)}_{\nu}(z) +\frac{\nu}{z} H^{(i)}_{\nu}(z)\,.
\label{Hankelprops4}
\end{eqnarray}
\end{subequations}

\section{Fourier Transforming Hankel Functions}
\label{Fourier Transforming Hankel Functions}

Consider the following integral equation:
\begin{equation}\label{I1}
I \equiv \int_{0}^{\infty} \mathrm{d}x \, x^{\mu+1} J_{\mu}(c x)
K_{\nu} (a x) K_{\nu} (b x)\,.
\end{equation}
Here, $ J_{\mu}(z)$ is a Bessel function of the first kind and
$K_{\nu} (z)$ is a MacDonald function. Note that for all integrals
in (\ref{Fourierintegral}) both $a$ and $b$ are purely imaginary.
The requirements from \cite{Gradshteyn} are such that:
\begin{subequations}
\label{I2}
\begin{eqnarray}
\mathrm{Re}(a) &>& 0 \label{I2a}\\
\mathrm{Re}(b) &>& 0 \label{I2b}\\
c &>& 0 \label{I2c}\\
\mathrm{Re}(\mu \pm \nu) &>& -1 \label{I2d}\\
\mathrm{Re}(\mu) &>& -1 \label{I2e}\,.
\end{eqnarray}
\end{subequations}
Hence we introduce small real contributions in the arguments of
both MacDonald functions only to take these to zero again at the
end of the calculation. We thus write:
\begin{subequations}
\label{I3}
\begin{eqnarray}
a &=& \delta + i \alpha \label{I3a}\\
b &=& \delta + i \beta \label{I3b}\,.
\end{eqnarray}
\end{subequations}
We choose $\delta>0$ to make the integral convergent. We now use
equation (6.578.10) from \cite{Gradshteyn} to find:
\begin{equation}\label{I4}
I = \frac{\sqrt{\pi} c^{\mu}
\Gamma(\mu+\nu+1)\Gamma(\mu-\nu+1)}{2^{3/2} (ab)^{\mu+1}
(u^{2}-1)^{\frac{1}{2}\mu+\frac{1}{4}}}
P^{-\mu-\frac{1}{2}}_{\nu-\frac{1}{2}}(u)\,,
\end{equation}
where:
\begin{equation}\label{I5}
u = \frac{a^{2}+b^{2}+c^{2}}{2ab}\,.
\end{equation}
Here, $ P^{-\mu-\frac{1}{2}}_{\nu-\frac{1}{2}}(u)$ represents the
associated Legendre function which we can rewrite in terms of a
Gauss' hypergeometric function $\!\phantom{1}_{2}\!F_{1}$ by means
of equations (8.702) and (9.131.1) of \cite{Gradshteyn}:
\begin{eqnarray}\label{I6}
I &=& \frac{\sqrt{\pi} (c/2)^{\mu}
\Gamma(\mu+\nu+1)\Gamma(\mu-\nu+1)}{4 (ab)^{\mu+1}
\Gamma(\mu+3/2)}  \\
&& \times_{2}F_{1}\left (\mu+\nu+1 , \mu-\nu+1; \mu+\frac{3}{2} ;
\frac{1-u}{2}\right) \nonumber \,.
\end{eqnarray}
Upon taking $\delta \rightarrow 0$ we find:
\begin{equation}\label{I7}
\lim_{\delta\rightarrow 0} \frac{1}{(ab)^{\mu+1}} =
\frac{1}{|ab|^{\mu+1}}\exp[-i\pi
(\mu+1)\theta(\alpha\beta)\mathrm{sgn}(\alpha + \beta)] \,.
\end{equation}
Finally note that we can further simplify the argument of the
hypergeometric function in equation (\ref{I6}) to:
\begin{equation}\label{I8}
\frac{1-u}{2} = 1+\frac{c^{2}-(\alpha + \beta +
i\delta)^{2}}{4\alpha\beta} \,.
\end{equation}
Let us finally stress that we have introduced $\delta>0$ to make
the integral convergent. It is however of significant physical
importance for it is the same $\delta$ that dictates the
$\varepsilon$ pole prescription in the propagator (see equations
(\ref{fermionpropagator}--\ref{fermionpropagator2}) and
(\ref{y})). This should not come as a surprise: we observe the
same behaviour in e.g. Minkowski and de Sitter spacetimes.

\end{document}